\documentclass[aps, 
twocolumn,review,
groupedaddress ,bibnotes]{revtex4-2}

 \usepackage{amsthm}
 \theoremstyle{plain}

 \usepackage{graphicx}
\usepackage{mathrsfs}
\usepackage{amssymb}
\usepackage{amsmath}
\usepackage{amsfonts,bm}
\usepackage[  colorlinks=true,
    linkcolor=blue,
    citecolor=blue]{hyperref}

\usepackage{pgffor}
\foreach \x in {A,...,Z}{
  \expandafter\xdef\csname  b\x\endcsname{\noexpand\mathbb{\x}}
}
\foreach \x in {A,...,Z}{
  \expandafter\xdef\csname c\x\endcsname{\noexpand\mathcal{\x}}
}
\foreach \x in {A,...,Z}{
  \expandafter\xdef\csname s\x\endcsname{\noexpand\mathscr{\x}}
}
\foreach \x in {A,...,Z}{
  \expandafter\xdef\csname sf\x\endcsname{\noexpand\mathsf{\x}}
}
\foreach \x in {a,...,z}{
  \expandafter\xdef\csname sf\x\endcsname{\noexpand\mathsf{\x}}
}
\foreach \x in {A,...,Z}{
  \expandafter\xdef\csname  fk\x\endcsname{\noexpand\frak{\x}}
}
\foreach \x in {a,...,z}{
  \expandafter\xdef\csname  fk\x\endcsname{\noexpand\frak{\x}}
}

    \newcommand{\PE}{\text{PE}}

\newcommand{\be}{\begin{equation}}
\newcommand{\ee}{\end{equation}}
\newcommand{\bpm}{\begin{pmatrix}}
\newcommand{\epm}{\end{pmatrix}}

\newcommand{\beqn}{\begin{eqnarray}}
\newcommand{\eeqn}{\end{eqnarray}}

\newcommand{\wt}{\widetilde}
\newcommand{\p}{\partial}
\newcommand{\CT}{\mathrm{CT}}

\begin{document}

\title{An Infinite Family of  Non-Rational VOAs from Strongly Coupled 4d Higgsless SCFTs
}

\author{Hongliang Jiang}
 \email[]{jianghongliang@fudan.edu.cn}
\affiliation{Center for Mathematics and Interdisciplinary Sciences, Fudan University, Shanghai 200433,
China}
\affiliation{Shanghai Institute for Mathematics and Interdisciplinary Sciences (SIMIS), Shanghai 200433,
China}

\begin{abstract}
We study  an infinite family of strongly coupled four-dimensional $\mathcal N=2$ superconformal field theories (SCFTs) distinguished   by a trivial Higgs branch, known as the $(A_2,D_{3m+1})$ Argyre-Douglas theories. We propose that their associated vertex operator algebras (VOAs) are the doublet algebras $\mathcal A(4m+2)$, an infinite family of non-rational vertex operator superalgebras with a remarkably simple strong generating set of only three fields. We provide several highly nontrivial   checks of this proposal. In particular, we reproduce the four-dimensional conformal anomalies $a$ and $c$ from the VOA   and analytically prove the exact equality between the Schur index of the SCFT and the supercharacter of the VOA. A key ingredient is a diagonal conformal gauging realization of the $(A_2,D_{3m+1})$ theories in terms of two simpler building blocks, which makes the Schur-index computation tractable. 
Remarkably, we find that the resulting Schur index of the $(A_2,D_{3m+1})$ theory coincides with that of $\mathcal N=4$ $SU(2)$ Super-Yang-Mills theory, up to an overall prefactor and an appropriate identification of fugacities.
We also discuss a generalization to the two-parameter family $(A_{2s},D_{(2s+1)m+1})$, whose members likewise have trivial Higgs branches and admit diagonal-gauging realizations. A particularly interesting subfamily is $(A_{2s},D_{2s+2})$, for which the four-dimensional conformal anomalies coincide, $a=c$.
As a byproduct, we also show that the doublet VOA $\cA(4m+2)$ is not classically free for positive integer $m $, and we discuss its implications for the Schur and Macdonald indices. 
 Our results reveal a systematic connection between   Higgsless SCFTs, conformal gauging, and strongly finite but non-rational vertex operator algebras.
\end{abstract}

\maketitle

\section{Introduction}

Vertex operator algebras (VOAs) play a central role in both mathematics and theoretical physics. Among them, rational VOAs  \footnote{In the physics literature, including this Letter, a ``rational VOA'' typically refers to a strongly rational VOA in the mathematical sense, incorporating both semisimplicity and $C_2$-cofiniteness. } are by far the best studied and best understood. They provide the chiral building blocks of rational conformal field theories and possess a remarkably rigid representation theory:  their module categories form modular tensor categories. This establishes a deep connection between rational VOAs, rational conformal field theory, and three-dimensional topological quantum field theory, and has also made them important in the study of topological phases of matter.

Nevertheless, rational VOAs occupy only a small corner of the much broader landscape of vertex operator algebras. Many physically and mathematically important VOAs are non-rational, with representation theories that are typically far richer and less well understood. A particularly compelling source of such algebras is provided by the four-dimensional SCFT/VOA correspondence, which associates to every four-dimensional $\mathcal N=2$ superconformal field theory (SCFT) a protected two-dimensional chiral algebra, or VOA \cite{Beem:2013sza}. This correspondence has revealed a striking bridge between strongly coupled four-dimensional quantum field theories and two-dimensional chiral algebraic structures.

The precise relation between the landscape of four-dimensional $\mathcal N=2$  SCFTs and that of VOAs, however, remains largely mysterious. The correspondence is neither injective nor surjective in any naive sense, and a natural long-term question is to characterize precisely which VOAs can arise from four-dimensional SCFTs and what four-dimensional information is encoded in their algebraic and representation-theoretic properties. This is a formidable problem, not least because neither side of the correspondence is presently classified.

A powerful constraint on the SCFT/VOA correspondence comes from the geometry of the VOA. Chiral algebras arising from four-dimensional $\mathcal N=2$ SCFTs are expected to be quasi-lisse \cite{Arakawa:2016hkg}, with their associated varieties closely related to the Higgs branches of the corresponding four-dimensional theories \cite{Beem:2017ooy}. This suggests focusing on particularly simple corners of the correspondence. Rational VOAs   have trivial associated variety and therefore can arise only from SCFTs with trivial Higgs branch.   The converse, however, need not hold: a Higgsless SCFT may instead give rise to a strongly finite  but non-rational, and hence logarithmic, VOA.

It is precisely this less explored possibility that we investigate in this Letter. We focus on the infinite family of $(A_2,D_{3m+1})$ Argyres--Douglas (AD) theories with trivial Higgs branch, and propose that their chiral algebras are the doublet algebras $\cA(4m+2)$ \cite{Feigin:2007sp,feigin2008characters}. For $m=1$, this reduces to the known correspondence between the $(A_2,D_4)$ AD theory and the $\cA(6)$ VOA \cite{Buican:2016arp}.

We provide several highly nontrivial checks of this proposal. We match the four-dimensional conformal anomalies $a$ and $c$ with the corresponding VOA data and analytically prove that the Schur index coincides with the vacuum supercharacter of $\cA(4m+2)$. A key ingredient is a diagonal-gauging realization of the $(A_2,D_{3m+1})$ theories in terms of two simpler AD building blocks with computable Schur indices. Remarkably, we find that the Schur index of this family of theories coincides with that of $\mathcal N=4$ $SU(2)$ Super-Yang-Mills  theory, up to an overall prefactor and an appropriate identification of fugacities.

 As a concrete illustration, we analyze the $(A_2,D_7)$/$\cA(10)$ pair in detail and discuss the modular properties of its index/character.  

We further extend this construction to the two-parameter family $(A_{2s},D_{(2s+1)m+1})$ whose members all have trivial Higgs branches. These theories likewise possess conformal manifolds and admit diagonal-gauging descriptions in terms of AD building blocks. The case $s=1$ reduces to the family studied above, while the subfamily $m=1$, namely $(A_{2s},D_{2s+2})$, has the notable property $a=c$.

Our infinite families of examples reveal a systematic connection between Higgsless
SCFTs, diagonal gauging, and strongly finite but non-rational vertex operator algebras, as well as a potentially intriguing relation to $\cN=4$ Super-Yang-Mills  theory.

The rest of this Letter is organized as follows. In section~\ref{su2gaugign}, we review the relevant properties of the $(A_2,D_{3m+1})$ theories and present their diagonal-gauging realization. In section~\ref{ApVOA}, we discuss the doublet algebras $\cA(4m+2)$, with emphasis on their vacuum supercharacters and modular properties. In section~\ref{VOASCFT}, we formulate the proposed   correspondence between $(A_2,D_{3m+1})$ and $\cA(4m+2)$, and present the main consistency checks. In section~\ref{generalizas}, we introduce the two-parameter generalization $(A_{2s},D_{(2s+1)m+1})$. We conclude in section~\ref{conslusion} with open questions and future directions, while technical details are deferred to the appendices.

 \section{$(A_2,D_{3m+1})$ SCFT and gauging}\label{su2gaugign}
 
 Let us first introduce some basic properties of $\cT_m\equiv (A_2,D_{3m+1})$ on which we focus in this Letter.  They   belong to a more general class of   SCFTs, denoted by $(G,G')$ with $G,G'$ both being  simply laced Lie algebras of ADE type, which can be obtained from type IIB string theory compactified on certain Calabi-Yau three-fold \cite{Cecotti:2010fi}. 
 
For $(A_2,D_{3m+1})$ Argyres-Douglas theory, this is engineered by compactifying type IIB string on a non-compact Calabi-Yau 3-fold, defined via a   quasihomogeneous hypersurface with isolated singularity  in $\bC^4$ given by 
$
 W(x,y,z,w)
=
x^3+y^{3m}+yz^2+w^2=0.
\label{eq:hypersurface}
$
Such a  type IIB realization allows for the immediate computation of  Coulomb branch (CB) spectrum:
\beqn
\cS_{\mathrm{CB}}(\cT_m) &=&
\Big\{\frac{2(j+m)}{2m+1},\frac{2(j+m)}{2m+1},\frac{2(j+2m)}{2m+1} \Big|
j= 1,\ldots, m   \Big\} 
\nonumber\\&& \quad
 \bigcup \Big\{\frac{3m+1}{2m+1}\Big\}~,
\eeqn
where repeated entries are counted with multiplicity. Therefore, the complex dimension  of the CB, called the rank, is $3m+1$. 
It is also easy to see  that there is precisely one Coulomb branch operator of dimension 2, which signals that the theory has a conformal manifold. In contrast, the Higgs branch is trivial, and the theory has no continuous flavor symmetry.  The type IIB realization also allows us to compute the central charges:
\be\label{cTm4d}
a_{\cT_m}
=
\frac{96m^2+43m+5}{24(2m+1)},
\qquad
c_{\cT_m}
=
\frac{(3m+1)(8m+1)}{6(2m+1)}.
\ee

  \begin{figure}[h]\centering
     \includegraphics[width=0.5 \textwidth]{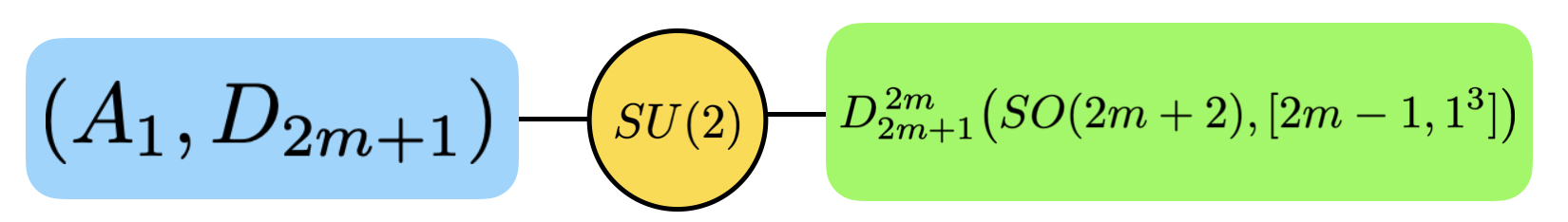}
     \caption{$(A_2,D_{3m+1})$ can be obtained from the diagonal gauging of $(A_1,D_{2m+1})$ and $D^{\,2m}_{2m+1}
\bigl(SO(2m+2) ,[2m-1,1^3] \bigr)$, both of which carry flavor symmetry $\fks\fku(2)$.  \label{ADgauge}}
\end{figure}

The presence of  an exactly marginal deformation   motivates the following gauging picture. 
The crucial point is that   the $(A_2,D_{3m+1})$ theory can be obtained from the diagonal conformal gauging of two theories as follows 
\footnote{This type of gauging realization was also observed in \cite{Carta:2021whq}. }:
\be\label{ADgauging}
\cT_m\equiv(A_2,D_{3m+1})
=
 {
(\mathfrak L_m\times\mathfrak R_m)
}/{
\fks\fku(2)_{\mathrm{diag}}
},
\ee
where
  \begin{align}\label{sfkLm}
\mathfrak L_m
&:= (A_1,D_{2m+1}),
\\
\mathfrak R_m
&:=
D^{\,2m}_{2m+1}
\bigl(SO(2m+2) ,[2m-1,1^3] \bigr) .
\end{align}
See Fig.~\ref{ADgauge} for illustration.  The two theories $\mathfrak L_m$ and $\mathfrak R_m$ are thus the building blocks, and both of them have $\fks\fku(2)$ flavor symmetry  \footnote{In this letter, we mainly focus on the local properties of the theories and do not care about their global forms, so we will not distinguish $SU(2)$ and $SO(3)$. }.
In the above formula, the theory for $\mathfrak R_m$ is written in the class-$\cS$   notation. In general, $D_p^b(G,Y)$   is constructed from the 6d (2,0) SCFT of type $G$ compactified on a sphere with two punctures: the irregular   puncture is determined by $p$ and $b$, and the regular puncture is specified by the nilpotent orbit $Y$ of $G$ \cite{Cecotti:2013lda,Xie:2012hs}.  For the full puncture, $Y$ is the trivial nilpotent orbit,  and its symbol can be omitted in the notation above. In the theory $\mathfrak R_m$ above, the irregular puncture does not contribute any flavor   symmetry and the specific choice of nilpotent orbit of the     regular puncture accounts for the entire residual       $\fks\fku(2)$  flavor symmetry.  

Note that in the special case of $m=1$, we have  $\mathfrak R_1=(A_1,D_3) \oplus (A_1,D_3)$ and the flavor symmetry is enhanced to $\fks\fko(4)\sim\fks\fku(2)\times \fks\fku(2)$. Furthermore \eqref{ADgauging} reduces to the diagonal gauging of three copies of $(A_1,D_3) $ theories. 

We now  justify the diagonal conformal gauging picture in \eqref{ADgauging}.

Using the results in appendix \ref{chargess}, one can verify that
\begin{align}\label{ack1}
a_{\mathfrak L_m}
+a_{\mathfrak R_m}
+a_{\mathrm{vec}}
&=
a_{\cT_m},
\\
c_{\mathfrak L_m}
+c_{\mathfrak R_m}
+c_{\mathrm{vec}}
&=
c_{\cT_m},
\\
 \cS_{\text{CB}}(\mathfrak L_m) \cup\cS_{\text{CB}}(\mathfrak R_m) \cup\cS_{\text{CB}}(\mathrm{vec})  
&=
\cS_{\text{CB}}(\cT_m).
 \end{align}
 Furthermore, one can also check that 
\be\label{ack2}
k_{\mathfrak L_m}^{4d}+k_{\mathfrak R_m}^{4d}=8=4h^\vee_{\fks\fku(2)} ,
\ee
which is precisely the anomaly free condition for conformal gauging.  All these relations provide strong evidence for the gauging realization in \eqref{ADgauging}.

 \section{$\cA(4m+2)$ VOA}\label{ApVOA}
  
  The doublet VOA $\cA(p)$  labelled by a positive integer $p$ was introduced in \cite{Feigin:2007sp,feigin2008characters}.  The central charge of the VOA is given by 
$
  c_{ \cA( p)} 
=13-6p- {6}/{p}
$. 
  %
  We will restrict to the case $p=4m+2$. 
This VOA has only   three strong generators, $T,\Psi,\widetilde\Psi$  with conformal weights $h_T=2,h_\Psi=h_{\wt\Psi}=\frac{3p-2}{4}=3m+1 $.  The first one is the stress tensor, and the latter two are fermionic primary operators. 
 The central charge is then
\be\label{cApcharge}
c_{ \cA( 4m+2)}  
=\frac{-2(3m+1)(8m+1)}{2m+1}. 
\ee
 
The OPEs $\Psi\times \Psi, \wt\Psi\times \wt\Psi$ 
are both regular, while $\Psi\times \wt\Psi$   is given by
\be
\Psi (z)\times \wt\Psi(w)=
\frac{1}{(z-w)^{2h_\Psi}}
\sum_{n= 0}^\infty(z-w)^n H_n(w),
\ee
where
\begin{equation}
H_0=\mathbf 1,
\quad
H_1=0,
\quad
H_2=\frac{2h_\Psi}{c_{\cA(p)}}T,
\quad
H_3=\frac{h_\Psi}{c_{\cA(p)}}\partial T, \quad \cdots.
\end{equation}
%
More generally, the $H_n$ are composite operators constructed from $T$ and its derivatives.

The defining feature of this VOA  is  the presence of the following null operators: 
\be\label{Tpsinull}
  \cN_\Psi:=\Psi''-2(2m+1)T\Psi  , \qquad   \cN_{\wt\Psi}:=\wt  \Psi''-2(2m+1)T\wt\Psi  .
\ee

  The VOA admits a free field realization as follows \cite{Feigin:2007sp,feigin2008characters}:  
  \beqn
 T  &=& \frac{1}{2} (\partial\varphi  )^2 + \frac{{4m+1}}{2  \sqrt{  {2m+1}}}\partial^2\varphi ,
 \\
\Psi   &=& e^{-\sqrt{2m+1}\varphi },  \label{Psiff}
\\
\wt \Psi (z) &=&\oint_{C_z} \frac{dz'}{2\pi i}     \, e^{2\sqrt{2m+1} \varphi(z')}\Psi (z)   \nonumber
\\&=&
 P_{4m+ 1}(\partial\varphi(z),\partial^2\varphi(z),\cdots) \, e^{\sqrt{2m+1}\varphi(z)},
\eeqn
where the free field obeys the OPE $\varphi(z)\varphi(w) \sim\log(z-w)$ and $P_{n }  $ is a   differential polynomial   of $\p^i \varphi$ with conformal weight $n$.

  The normalized character of   various modules $\cM_s$ of $\cA(p)$ VOA  is given in  \cite{Feigin:2007sp,feigin2008characters}:  
  \beqn
  \text{ch}_s(q,y)&=&\text{tr}_{\cM_s} q^{L_0} y^Q
  \nonumber \\ &=&
 \frac{q^{-\frac{(p-s)^2}{4p}}}{ (q;q)_\infty  } \sum_{n =1}^\infty \sum_{j = -\frac{n-1}{2}}^{\frac{n-1}{2}} y^{2 j} \left( q^{\frac{p}{4}(n - \frac{s}{p})^2} - q^{\frac{p}{4}(n + \frac{s}{p})^2} \right),  \nonumber\\
  \eeqn
  where $(q;q)_\infty =\prod_{n=1}^\infty (1 - q^n)$, and $Q$ assigns charge $\pm 1$ to $\Psi,\wt\Psi$,  respectively.
  
We are mainly interested in the   vacuum module which corresponds to $s=1$ above. Furthermore, we set $y=-1$ which yields the supercharacter  of the  $\cA(4m+2)$ VOA
\beqn 
&&\chi_{\cA(4m+2)}(q) =
\text{str } q^{L_0}=\text{ch}_1(q,-1)
\\&=& \label{superch}
\frac{1}{(q;q)_\infty}
\sum_{r=1}^{\infty}
(-1)^{r-1}r\,q^{\frac{(r-1)((2m+1)r+2m)}{2}}
(1-q^r) .
\qquad
\eeqn

   \section{The correspondence between $(A_2,D_{3m+1})$ SCFT and $\cA(4m+2)$ VOA }\label{VOASCFT}
     In this section, we provide compelling  evidences from multiple  perspectives that the chiral algebra of  $(A_2,D_{3m+1})$ SCFT is    the $\cA(4m+2)$ VOA.  
   \subsection{Central charges}
  By  comparing   \eqref{cApcharge}
and  \eqref{cTm4d}, it is easy to see that $c_{ \cA( 4m+2)} =-12 c_{\cT_m} $,  which provides the   first evidence for the proposed correspondence \cite{Beem:2013sza}. Next, we show that  $a$-central charge inferred from the VOA data  also agrees with the 4d SCFT.

   Using Poisson  resummation, one can show that the supercharacter \eqref{superch} can be rewritten as
   \beqn
\chi_{\cA(4m+2)} &=&
   \frac{q^{ -m}}{(q;q)_\infty}
 \frac{ i}{ ( (2m+1) \tau/i )^{3/2}}
 \nonumber\\&& \times \sum_{n \in \mathbb{Z}} (n-\frac{1-\tau}{2}) e^{-\frac{\pi i (n-\frac{1-\tau}{2})^2}{(2m+1) \tau}  }~.
   \eeqn
   In the high temperature limit $\tau\to 0$, the dominant contribution comes from the $n=0$ and $n=1$ terms: 
   \be\label{chihT}
 \chi_{\cA(4m+2)}= \frac{ i }{(2m+1)^{3/2} \tau }     \sin  \frac{\pi  }{4m+2  }   e^{ \frac{\pi i (m-1)}{6 (2m+1) \tau}    } +\cdots,
   \ee
 where the ellipsis represents subleading terms in the limit $\tau \to 0$. 
 The high temperature limit  is controlled by the effective central charge 
 $\lim_{\tau\to 0} \log \chi =\frac{\pi i}{12\tau}c_\text{eff}+\cdots $ \cite{Beem:2017ooy}, which implies 
  \be
c_\text{eff}=\frac{2(m-1)}{   2m+1  } .
\ee
On the other hand, it is known that the 2d effective central charge is related to the 4d central charges   $c_\text{eff}= 48(c_{4d}-a_{4d})$ \cite{DiPietro:2014bca,Buican:2015ina}. 
 Using this relation, one can  readily compute the  $a$ central charge, which is exactly the value given  in \eqref{cTm4d}. This provides the   second evidence for the proposed correspondence.

\subsection{Schur index}
 Now we provide a third, and the strongest, check by matching the character of VOA and index of SCFT, which is manageable thanks to the gauging realization~\eqref{ADgauging}.

First, the Schur index of $\mathfrak L_m$  is \cite{Song:2017oew}
\be
\mathcal I^S_{\mathfrak L_m}(q,z)
=
\PE\left[
\frac{q-q^{2m+1}}
{(1-q)(1-q^{2m+1})}
\chi_{\mathbf 3}(z)
\right],
 \label{eq:left-index}
\ee
where $\PE$ is the plethystic exponential, and $\chi_{\mathbf 3}(z)=z^2+1+z^{-2}$ is the character of the adjoint representation of $\fks\fku(2)$ flavor symmetry and $z$ is the corresponding fugacity.

Next, we need to compute the   Schur index of $\mathfrak R_m$, which requires a deeper understanding of the regular puncture and the corresponding nilpotent orbit. 
The nilpotent orbit $ [2m-1,1^3]$ induces the decomposition $\fks\fko({2m+2})\supset \fks\fku(2)_{X} \times \fks\fku(2)_f $, where the former   denotes the embedding $\fks\fku(2)$,  and the latter denotes the   residual flavor $\fks\fko(3)\sim\fks\fku(2)$, namely the commutant of $  \fks\fku(2)_X$ in  $\fks\fko({2m+2})$. 
Consequently the adjoint representation of $\fks\fko({2m+2})$, which is just the antisymmetric tensor product of two vector representations, decomposes under $   \fks\fku(2)_f  \times  \fks\fku(2)_{X} $ as
\begin{equation}\label{sodecmp}
\Lambda^2({\bm{2m+2}})
\cong
\Big[\bigoplus_{a=1}^{m-1}
( \mathbf 1 \otimes V_{2a-1}  ) \Big]
\oplus
( \mathbf 3 \otimes V_{m-1}  )
\oplus
(\mathbf 3\otimes V_0  ),
\end{equation}
where \(V_j\) is the spin-\(j\) irreducible representation under $\fks\fku(2)_{X}$, and boldface number   $\bm n$ on the RHS denotes the $n$-dimensional irreducible representation of  flavor $ \fks\fku(2)_f $. 

With \eqref{sodecmp}, the Schur index of $\fkR_m$ 
  can be computed just by applying the formula in \cite{Xie:2019zlb}  
\beqn
&&\mathcal I^S_{\mathfrak R_m}(q,z)
\nonumber \\   &=&
\PE\left[
\frac{
(q^2-q^{2m})
+
(q-q^{2m+1}+q^m-q^{m+2})
\chi_{\mathbf 3}(z)
}{
(1-q)(1-q^{2m+1})
}
\right].
\nonumber \\
\label{eq:right-index}
\eeqn

The gauging procedure \eqref{ADgauging} then allows us to compute  the
Schur index of $\cT_m$ as follows:
\beqn
&&\mathcal I^S_{\cT_m}(q) 
=
\int d\mu_{\fks\fku(2)}(z)\,
\cI^S_\text{vec} \cI^S_{\mathfrak L_m} \cI^S_{\mathfrak R_m}
\nonumber \\ &=&
\frac12 
\PE\Big[ 
\frac{(q^2-q^{2m})
}{
(1-q)(1-q^{2m+1})}
\Big]
\oint\frac{dz}{2\pi i z}(1-z^2)(1-z^{-2})
\nonumber \\&&
\times\PE\Big[ 
\frac{ 
q^m(1+q-2 q^{m+1})
\chi_{\mathbf 3}(z)
}{
 (1-q^{2m+1})
}
\Big] , 
\label{IschurTm}
\eeqn
where  $\mathcal I^S_{\mathrm{vec}}(q,z) 
= \PE [-\frac{2q}{1-q}\chi_{\mathbf 3}(z) ]$ is    the vector multiplet contribution.
 
 This can be compared with the Schur index of $\cN=4 $ SU(2) Super-Yang-Mills (SYM) theory
\beqn
&&
 \cI^S_ { \cN=4    \text{ SYM}}(q, y)
 =\int d\mu_{\fks\fku(2)}(z) \, \cI^S_\text{hyp}  \cI^S_\text{vec} 
\nonumber \\&=&
 \int d\mu_{\fks\fku(2)}(z)  \, \text{PE} \left[ \left(  \frac{-2q}{1-q} + \frac{q^{\frac{1}{2}}(y+y^{-1}) }{1-q}  \right) \chi_{\bf 3} (z) \right],
 \qquad
\eeqn
where $\cI^S_{\text{hyper}}(q, y, z) = \PE [ \frac{q^{\frac{1}{2}}}{1-q}  (y+\frac{1}{y} ) \chi_{\bf 3} (z)  ]$, and $y$ is the fugacity of the $SU(2)_F\subset SU(4)_R$   of $\cN = 4$ SYM.

An easy comparison shows that 
\beqn
\mathcal I^S_{\cT_m}(q) &=&\PE\Big[  \frac{(q^2-q^{2m}) }{ (1-q)(1-q^{2m+1})} \Big]
 \cI^S_ { \cN=4    \text{ SYM}}(q^{2m+1}, q^{\frac12})
\nonumber \\&=&
  \cI_{(A_1,A_{2m-2})} (q)
 \cI^S_ { \cN=4    \text{ SYM}}(q^{2m+1}, q^{\frac12}) .
\eeqn
This generalizes the  special  case of $m=1$ in \cite{Buican:2020moo}, where the overall prefactor is trivial.  

Our central claim is that  the supercharacter  of VOA \eqref{superch} is equal to the Schur index of SCFT  \eqref{IschurTm}, namely:
\be\label{ISischi}
\chi_{\cA(4m+2)}(q)=\mathcal I^S_{\cT_m}(q).
\ee
For small values of $m$, both sides can be computed numerically to very high orders. Indeed they agree with each other and  are given by 
\begin{align}
m = 1:&\; 1 + q^2 + q^3 + 2q^6 + q^8 + q^{11} + 2q^{12}  +\cdots,
\label{A4Schur}
\\
m = 2:&\; 1 + q^2 + q^3 + 2q^4 + 2q^5 + 4q^6 + 2q^7 + 5q^8 
 \nonumber  \\ &
 \;  + 6q^9  + 8q^{10} + 8q^{11} + 13q^{12} +\cdots , 
\\
m = 3:&\; 1 + q^2 + q^3 + 2q^4 + 2q^5 + 4q^6 + 4q^7 + 7q^8
 \nonumber  \\ &
 \; + 8q^9 
+ 10q^{10} + 12q^{11} + 19q^{12}+\cdots.  
\end{align}

 In fact, the equality in \eqref{ISischi} can be shown exactly. We prove this in the appendix~\ref{ISischiapp}.

One can also use a similar  gauging strategy   to compute the Hall–Littlewood (HL) indices. By explicit computation, we find that  the HL index  is also  trivial $\cI_{\cT_m}^\text{HL}=1$. This is consistent with the fact that the theory  $\cT_m$ has no Higgs branch. Indeed, since $h+R-r\in \bZ$ for Schur operators  and $\Psi,\wt\Psi$ are fermionic, we thus have $R,r\in \bZ+\frac12$ and $h\in \bZ$. Therefore $\Psi,\bar\Psi$ can not be $\hat\cB$ type Higgs branch operator, which requires $h=R$ and $r=0$.

The equality between  index and character implies the agreement of an infinite amount of characteristic data, thereby providing very  compelling   evidence that the associated VOA of $\cT_m$ is indeed given by $\cA(4m+2)$.
 
 \subsection{Modularity}
 Now, we study the modularity of the VOA. 
 For $m=1$, the modular properties  have been studied in \cite{Jiang:2024baj}. So we consider the next simplest example with $m=2$. 
 For $m=2$,  we have null relations $\Psi''=10T\Psi, \wt\Psi''=10T\wt \Psi$. We also find a null relation at weight 16 of the form $T^8=\cdots$, where   every term in  the ellipsis contains at least one derivative, namely $T^8\in C_2(\cV)$. This implies that the  modularly normalized vacuum character/index satisfies an order-8 modular linear differential equation (MLDE) whose explicit form is found to be
 \beqn
&&
\Big[ D_q^{(8)} - \frac{3948}{5} \bE_4 D_q^{(6)} - \frac{44856}{5} \bE_6 D_q^{(5)} 
  + \frac{844518}{25} \bE_4^2 D_q^{(4)} 
 \nonumber   \\&&
    + \frac{10480848}{25} \bE_4 \bE_6 D_q^{(3)}     
    + \frac{(333085204 \bE_4^3 - 315932400 \bE_6^2)}{125} D_q^{(2)} 
     \nonumber   \\&&   
 +   \frac{52466568}{125} \bE_4^2 \bE_6 D_q^{(1)}
    \nonumber   \\&& 
        + \frac{(-377011635 \bE_4^4 + 1537331488 \bE_4 \bE_6^2)}{125}  
        \Big]  \cZ_{\cT_2} =0 ,\qquad\qquad\quad
\eeqn
where $\cZ_{\cT_2}=q^{-c_{2d}/24}\cI^S_{\cT_2}=q^{119/60}\cI^S_{\cT_2}$. 
  The solution to the indicial equation of  the MLDE yields the following distinct conformal weights of the ordinary modules
  $h=-2,-  \frac{9}{5},-\frac{7}{5} ,-\frac{4}{5},0$, with the first three appearing as double roots which yield extra logarithmic solutions.
  The solutions to MLDE, namely the (generalized) characters,  and the modular $S,T$ matrices are given in appendix~\ref{mST}.

 We conjecture that in general the MLDE  for $\cT_m$ has order   $3m+2$, which has been verified for $m=1,2,3$.

 \subsection{Classical non-freeness and Macdonald index}

We now show that the doublet VOA $\cA(4m+2)$
is not classically free for $m\geq 1$.

Recall that a vertex operator (super)algebra $\cV$ is called
\emph{classically free} if the canonical surjective homomorphism
\be\label{mapf}
f: J_\infty(R_\cV)\twoheadrightarrow \operatorname{gr}^{F}\cV
\ee
is injective, and hence an isomorphism. Here $R_\cV=\cV/C_2(\cV)$
is Zhu's $C_2$-algebra, $J_\infty(R_\cV)$ is its differential
supercommutative arc algebra, and $\operatorname{gr}^{F}\cV$ is
the associated graded algebra with respect to Li's filtration.
Geometrically, $J_\infty(R_\cV)$ is the coordinate algebra of the
corresponding arc superscheme or infinite jet superscheme. See \cite{ArakawaMoreau2021} for details. 

In the present case, the  Zhu's $C_2$-algebra is
\be\label{Ram}
R_{\cA(4m+2)}
=
\frac{\bC[t]\otimes\Lambda(\xi,\eta)}
{(t\xi,\ t\eta,\ \xi\eta-\kappa_m t^{3m+1})},
\ee
where $\xi=[\Psi]$ and $\eta=[\widetilde{\Psi}]$ are odd, while
$t=[T]$ is even. The coefficient $\kappa_m$ is nonzero and can
be normalized to one by rescaling the odd generators.
The quotient relations follow from \eqref{Tpsinull} and the
null relation of conformal weight $6m+2$; see
\cite{Jiang:2024baj} for the explicit form of the latter at $m=1$.

Let $D$ denote the even derivation of the arc algebra.
We claim that
\be
0\neq \xi D\xi\in\ker f.
\ee
To establish nontriviality of the kernel, consider the quotient
$
R_{\cA(4m+2)}/(t,\eta)\cong\Lambda(\xi).
$
It induces a surjective homomorphism
\be
J_\infty\bigl(R_{\cA(4m+2)}\bigr)
\twoheadrightarrow
J_\infty\bigl(\Lambda(\xi)\bigr)
=
\Lambda(\xi,D\xi,D^2\xi,\ldots).
\ee
The image of $\xi D\xi$ is the product of two distinct odd
generators and is therefore nonzero. Notice that
$
D(\xi^2)=(D\xi)\xi+\xi D\xi=0
$
holds identically by supercommutativity and does not imply
$\xi D\xi=0$.

On the other hand, the free-field realization \eqref{Psiff}
gives the self-OPE
$
\Psi(z)\Psi(0)=\mathcal{O}(z^{2m+1})
$
\cite{Feigin:2007sp,feigin2008characters}. Consequently,
\be
:\!\Psi\,\p^k\Psi\!:\,=0,
\qquad k=0,\ldots,2m.
\ee
Since $m\geq1$, this includes the null relation
$:\!\Psi\,\p\Psi\!:\,=0$, which implies $f(\xi D\xi) =0$.
Thus
$
0\neq\xi D\xi\in\ker f,
$
proving that $\cA(4m+2)$ is not classically free.  This provides a  family of concrete counterexamples to the conjecture that all VOAs arising from 4d SCFTs are classically free. 

If the canonical map $f$ in \eqref{mapf} is an isomorphism, the conformally graded  super Hilbert   series of the
arc algebra must coincide with the normalized vacuum supercharacter
of the VOA, and hence with the Schur index of the associated
four-dimensional SCFT.  

Consider first $m=1$, corresponding to $\mathcal A(6)$. Assigning
the conformal weights $ \deg_q(t,\xi,\eta,D)=(2,4,4,1)$, 
   we obtain
\begin{equation}
\mathrm{HS}(q)
=
1+q^2+q^3+2q^6+q^8+3q^9+3q^{10}
+4q^{11}+5q^{12}+O(q^{13}).
\end{equation}
This differs from \eqref{A4Schur} already at order $q^9$.
The mismatch proves that $f$ is not an isomorphism and that
$\mathcal A(6)$ is not classically free.  By contrast, the $(A_1,A_{2n})$ AD
theories are associated with the classically free $(2,2n+3)$
Virasoro minimal models, whose arc-space Hilbert series reproduce
their Schur   indices  \cite{Bhargava:2023hsc}.

However, if we impose three additional relations
$
\xi D\xi=0, 
\eta D\eta=0, 
\xi D\eta-(D\xi)\eta=0,
$
together with all their derivatives in $ {J_\infty(R_{\mathcal A(6)})}$,  we get Hilbert   series  
\begin{align}
\mathrm{HS}'(q)
={}&1+q^2+q^3+2q^6+q^8+q^{11}+2q^{12}+q^{15}
+2q^{18}\nonumber\\
&\quad
+2q^{20}+q^{24}+q^{26}+2q^{27}+2q^{30}
+O(q^{31}),
\end{align}
in agreement with \eqref{IschurTm}.

We can further assign the second   degree  $
\deg_T(t,\xi,\eta,D)=(1,2,3,0).
$
This is a filtration rather than a grading, since the
$C_2$ relation $\xi\eta\propto t^4$ is not homogeneous in these
degrees. The bigraded Hilbert super series  can be computed
\begin{align}
&\mathrm{HS}'(q,T)
\nonumber \\={}&1+Tq^2+Tq^3+(T-T^3)(q^4+q^5)
+(T+T^2)q^6\nonumber\\
&+(T+T^2-T^3-T^4)q^7
+(T+2T^2-T^3-T^4)q^8\nonumber\\
&+(T+2T^2-T^3-2T^4)q^9
+(T+3T^2-T^3-3T^4)q^{10}\nonumber\\
&+(T+3T^2-T^3-3T^4+T^6)q^{11}\nonumber\\
&+(T+4T^2-4T^4+T^6)q^{12}
+O(q^{13}),
\end{align}
which agrees with the Macdonald index in \cite{Jiang:2026ffv}.

More generally, consider the differential ideal
\begin{equation}
\mathcal K_m
=
\left\langle
\xi D^{2j-1}\xi,\,
\eta D^{2j-1}\eta,\,
\xi D^{2j-1}\eta-(D^{2j-1}\xi)\eta
\; : \;
1\le j\le m
\right\rangle_D
\end{equation}
and the quotient $B_m=
J_\infty\!\left(R_{\mathcal A(4m+2)}\right)/\mathcal K_m.
$
We find that the corresponding Hilbert   series  agrees with the
Schur index. We have verified this agreement for $m=1,\ldots,10$
through order $q^{80}$. %

To extend this construction to the Macdonald index for \(m>1\),
we propose a refinement of the filtration by
introducing redundant even differential generators \(u_j\) as follows:
\[
\widetilde{B}_m = \frac{B_m[u_j, Du_j, \cdots]_{1 \leqslant j < m}}{\langle D^k(u_j - \xi D^{2j }\eta) : k \geqslant 0, 1 \leqslant j < m \rangle} .
\] 
This refinement is motivated by the null relations in VOA $\Psi\p^{2j}\widetilde \Psi =c_j T^{j+3m+1}+ \text{terms with fewer } T$ where $c_j\neq 0$.  
Eliminating the \(u_j\) by setting \(u_j=\xi D^{2j}\eta\) gives an
isomorphism
$
\widetilde B_m\simeq B_m
$
of differential superalgebras.  This however does not hold under the $T$-filtration below: 

\begin{center}
\renewcommand{\arraystretch}{1.5}
\begin{tabular}{c|c|c|c}
\hline
generator & $q$-weight &  $T$-weight & parity \\
\hline
\(t\) & \(2\) & \(1\) & even \\
\(\xi\) & \(3m+1\) & \(2m\) & odd \\
\(\eta\) & \(3m+1\) & \(2m+1\) & odd \\
\(u_j\) & \(2(3m+1+j)\) & \(3m+1+j\) & even \\
\(D\) & \(1\) & \(0\) & even \\
\hline
\end{tabular}
\end{center}

We propose that the  bigraded Hilbert series of $\widetilde{B}_m $ gives the Macdonald index. 
These will be developed in detail in~
\cite{Jiang2026}, where we will  put forward  the following closed expression for Macdonald index \footnote{For $m=1$, this expression agrees with the proposal of Ref.~\cite{Yoshida:2026ugm}.}
\begin{equation}\label{IM22}
\mathcal I_M 
=
\sum_{\bm n\ge0}
\frac{
(-1)^{n_++n_-}
q^{E(\bm n)}
T^{\,2mn_++(2m+1)n_-+\sum_{i=1}^{4m+1}i n_i}
}{
(q;q)_{n_+}(q;q)_{n_-}
\prod_{i=1}^{4m+1}(q;q)_{n_i}
},
\end{equation}
where
$
\bm n=(n_+,n_-,n_1,\ldots,n_{4m+1})
$
and
\begin{equation}
E(\bm n)
=
\sum_A d_A n_A
+
\sum_A K_{AA}\binom{n_A}{2}
+
\sum_{A<B}K_{AB}n_A n_B.
\end{equation}
Here $A,B\in\{+,-,1,\ldots,4m+1\}$, $
d_+=d_-=3m+1,  d_i=2i,  $
and the symmetric matrix $K$ is given by
\begin{equation}
K_{\pm\pm}=K_{\pm\mp}=2m+1,\quad
K_{\pm,i}=i,\quad
K_{ij}=2\min(i,j).
\end{equation}

The proposed sum \eqref{IM22} has also  an exact algebraic realization as the
bigraded Hilbert super series of  
\begin{equation}
\mathcal Q_K
=
\frac{
\mathbb C[D^rX_i\mid r\ge0 ]
\otimes
\Lambda(D^rX_+,D^rX_-\mid r\ge0)
}{
\left\langle
D^\ell\!\left(X_A D^sX_B\right)
\;\middle|\;
\ell\ge0,\quad 0\le s<K_{AB}
\right\rangle
}.
\end{equation}
where  generators $X_i$ are even and $X_\pm$ are odd, with
\begin{align}
\deg_{q,T}X_i&=(2i,i),
&&1\le i\le4m+1,\nonumber\\
\deg_{q,T}X_+&=(3m+1,2m),&
\deg_{q,T}X_-&=(3m+1,2m+1),\nonumber\\
\deg_{q,T}D&=(1,0).
\end{align}
Setting $T=1$ recovers the prescription of  computing VOA  vacuum supercharacter, or equivalently the Schur index,  in \cite{Feigin:2007sp,feigin2008characters}.

 \section{Generalization}\label{generalizas}

  We now generalize our previous examples to a two-parameter family of theories. It is described by    
 \be\label{Tms}
\cT_{m,s}\equiv\left( A_{2s}, D_{(2s+1)m+1} \right)
=( \fkL_m  \times\fkR_{m,s}  )/\fks\fku(2)_{\mathrm{diag}},
\ee
where $\fkL_m$ is given in \eqref{sfkLm} and 
\beqn
\fkR_{m,s}&=&  D_{(2s-1)(2m+1)}^{2(2s-1)m} \bigl( SO(2(2s-1)m+2), 
\nonumber\\ &&\qquad\qquad\qquad\qquad
[ 2(2s-1)m-1,1^3 ] \bigr) .\qquad
\eeqn
 This family of theories $\cT_{m,s}$ likewise have trivial Higgs branches. The gauging \eqref{Tms} can be similarly justified by verifying the   same relations as in \eqref{ack1}-\eqref{ack2}.  In particular, we provide the central charges of $\fkR_{m,s}$ in appendix \ref{chargess}. 
 
If $s=1$, we recover the  $(A_{2}, D_{ 3m+1})$ studied before. 
   If $m=1$, we get $(A_{2s}, D_{ 2s+2})$. A notable property of this family of theories is that the two central charges coincide,  $a=c=\frac{1}{9} s (s+1) (4s+5)$. The     
   rank $r_\text{CB}=2s(s+1)$, and  the theory has a conformal manifold of dimension $2s-\delta_{s,1}$. 
 This feature generalizes the    $(A_2,D_4) $  theory corresponding to $m=s=1$.
   
The associated VOAs of these two-parameter families are more intricate and will be studied in \cite{Jiang2026}.

\section{Conclusion}\label{conslusion}

In this letter, we have  proposed the    correspondence between the $(A_2,D_{3m+1})$ SCFTs and the doublet algebras $\cA(4m+2)$. The proposal is substantiated from multiple perspectives, including agreement of infinitely many characteristic data. A natural next question concerns the closely related family $\cA(4m)$, for which all three strong generators are bosonic. It has been proposed that the simplest example, $\cA(4)$, is associated with a certain rank-one SCFT which also has no Higgs branch \cite{Deb:2025cqr}. It would be interesting to scrutinize this proposal further and to determine the 4d SCFT origins of the full family of   $\cA(4m)$ VOAs.

Another important direction is to identify the VOAs associated with the more general theories
$ (A_{2s},D_{(2s+1)m+1})$ introduced above. Of particular interest is the subfamily with $m=1$, for which $a=c$. A major obstacle is that a direct computation of the Schur index becomes considerably more subtle, since the relevant building blocks involve non-admissible affine levels. Developing alternative methods to determine their Schur indices and chiral algebras would therefore be highly desirable. Details will be discussed in \cite{Jiang2026}.

The examples studied in this paper arise from the diagonal gauging of two building blocks. It is natural to ask what happens when three or more building blocks are involved. A particularly interesting class is provided by the theories $\cT_{(p,N)}$~\cite{Jiang:2024baj}, labelled by two coprime integers with $p=2,3,4,6$, which enjoy the special property $a=c$. Among them, $\cT_{(3,2)}$, $\cT_{(4,3)}$, and $\cT_{(6,5)}$ are Higgsless. The first, $\cT_{(3,2)}$, can be identified with the $(A_2,D_4)$ theory corresponding to  $\cA(6)$  chiral algebra, whereas the VOAs associated with $\cT_{(4,3)}$ and $\cT_{(6,5)}$ remain unknown and would be particularly interesting to determine.

More generally, one expects many further examples of Higgsless SCFTs whose associated VOAs are nevertheless non-rational. In the Lagrangian setting, Higgsless SCFTs were classified in Ref.~\cite{Chen:2026vyz}, where, in particular, one infinite families were identified, although their associated VOAs remain unknown. It is therefore natural to ask whether a similar classification can be achieved for Higgsless SCFTs constructed from non-Lagrangian building blocks. All these examples     suggest that conformal gauging may play a central role in producing Higgsless SCFTs with non-rational chiral algebras. Clarifying the precise relation between conformal gauging, the absence of a Higgs branch, and non-rationality of the associated VOA would be an important direction for future work. Motivated by known examples, we conjecture that, for SCFTs with trivial Higgs branch, the  associated VOA is strongly  rational if all the  Coulomb dimensions are non-integral, and strongly finite but non-rational otherwise.

Finally, it would be valuable to develop a systematic understanding of the representation theory of these non-rational VOAs. In particular, one would like to classify their irreducible and indecomposable modules, determine their modular properties, and understand how these representation-theoretic structures are encoded in the corresponding four-dimensional SCFTs.
 
 \bigskip

 \noindent Note added: After the submission of the first version of this paper to arXiv, Ref.~\cite{Yoshida:2026ugm} appeared, which has some overlap with this Letter.

 \bigskip
 \noindent
\emph{Acknowledgement.}
The author would like to thank  Leonardo Rastelli for   correspondence that inspired this work. 
This work was supported by the startup
grant at SIMIS and the Shanghai Pujiang Program (No. 25PJA128).

 \bigskip \bigskip
\hfill 

 \bibliography{ref}

\begin{thebibliography}{29}%
\makeatletter
\providecommand \@ifxundefined [1]{%
 \@ifx{#1\undefined}
}%
\providecommand \@ifnum [1]{%
 \ifnum #1\expandafter \@firstoftwo
 \else \expandafter \@secondoftwo
 \fi
}%
\providecommand \@ifx [1]{%
 \ifx #1\expandafter \@firstoftwo
 \else \expandafter \@secondoftwo
 \fi
}%
\providecommand \natexlab [1]{#1}%
\providecommand \enquote  [1]{``#1''}%
\providecommand \bibnamefont  [1]{#1}%
\providecommand \bibfnamefont [1]{#1}%
\providecommand \citenamefont [1]{#1}%
\providecommand \href@noop [0]{\@secondoftwo}%
\providecommand \href [0]{\begingroup \@sanitize@url \@href}%
\providecommand \@href[1]{\@@startlink{#1}\@@href}%
\providecommand \@@href[1]{\endgroup#1\@@endlink}%
\providecommand \@sanitize@url [0]{\catcode `\\12\catcode `\$12\catcode
  `\&12\catcode `\#12\catcode `\^12\catcode `\_12\catcode `\%12\relax}%
\providecommand \@@startlink[1]{}%
\providecommand \@@endlink[0]{}%
\providecommand \url  [0]{\begingroup\@sanitize@url \@url }%
\providecommand \@url [1]{\endgroup\@href {#1}{\urlprefix }}%
\providecommand \urlprefix  [0]{URL }%
\providecommand \Eprint [0]{\href }%
\providecommand \doibase [0]{https://doi.org/}%
\providecommand \selectlanguage [0]{\@gobble}%
\providecommand \bibinfo  [0]{\@secondoftwo}%
\providecommand \bibfield  [0]{\@secondoftwo}%
\providecommand \translation [1]{[#1]}%
\providecommand \BibitemOpen [0]{}%
\providecommand \bibitemStop [0]{}%
\providecommand \bibitemNoStop [0]{.\EOS\space}%
\providecommand \EOS [0]{\spacefactor3000\relax}%
\providecommand \BibitemShut  [1]{\csname bibitem#1\endcsname}%
\let\auto@bib@innerbib\@empty
\bibitem [{Note1()}]{Note1}%
  \BibitemOpen
  \bibinfo {note} {In the physics literature, including this Letter, a
  ``rational VOA'' typically refers to a strongly rational VOA in the
  mathematical sense, incorporating both semisimplicity and
  $C_2$-cofiniteness.}\BibitemShut {Stop}%
\bibitem [{\citenamefont {Beem}\ \emph {et~al.}(2015)\citenamefont {Beem},
  \citenamefont {Lemos}, \citenamefont {Liendo}, \citenamefont {Peelaers},
  \citenamefont {Rastelli},\ and\ \citenamefont {van Rees}}]{Beem:2013sza}%
  \BibitemOpen
  \bibfield  {author} {\bibinfo {author} {\bibfnamefont {C.}~\bibnamefont
  {Beem}}, \bibinfo {author} {\bibfnamefont {M.}~\bibnamefont {Lemos}},
  \bibinfo {author} {\bibfnamefont {P.}~\bibnamefont {Liendo}}, \bibinfo
  {author} {\bibfnamefont {W.}~\bibnamefont {Peelaers}}, \bibinfo {author}
  {\bibfnamefont {L.}~\bibnamefont {Rastelli}},\ and\ \bibinfo {author}
  {\bibfnamefont {B.~C.}\ \bibnamefont {van Rees}},\ }\bibfield  {title}
  {\bibinfo {title} {{Infinite Chiral Symmetry in Four Dimensions}},\ }\href
  {https://doi.org/10.1007/s00220-014-2272-x} {\bibfield  {journal} {\bibinfo
  {journal} {Commun. Math. Phys.}\ }\textbf {\bibinfo {volume} {336}},\
  \bibinfo {pages} {1359} (\bibinfo {year} {2015})},\ \Eprint
  {https://arxiv.org/abs/1312.5344} {arXiv:1312.5344 [hep-th]} \BibitemShut
  {NoStop}%
\bibitem [{\citenamefont {Arakawa}\ and\ \citenamefont
  {Kawasetsu}(2016)}]{Arakawa:2016hkg}%
  \BibitemOpen
  \bibfield  {author} {\bibinfo {author} {\bibfnamefont {T.}~\bibnamefont
  {Arakawa}}\ and\ \bibinfo {author} {\bibfnamefont {K.}~\bibnamefont
  {Kawasetsu}},\ }\bibfield  {title} {\bibinfo {title} {{Quasi-lisse vertex
  algebras and modular linear differential equations}},\ }\href@noop {} {\
  (\bibinfo {year} {2016})},\ \Eprint {https://arxiv.org/abs/1610.05865}
  {arXiv:1610.05865 [math.QA]} \BibitemShut {NoStop}%
\bibitem [{\citenamefont {Beem}\ and\ \citenamefont
  {Rastelli}(2018)}]{Beem:2017ooy}%
  \BibitemOpen
  \bibfield  {author} {\bibinfo {author} {\bibfnamefont {C.}~\bibnamefont
  {Beem}}\ and\ \bibinfo {author} {\bibfnamefont {L.}~\bibnamefont
  {Rastelli}},\ }\bibfield  {title} {\bibinfo {title} {{Vertex operator
  algebras, Higgs branches, and modular differential equations}},\ }\href
  {https://doi.org/10.1007/JHEP08(2018)114} {\bibfield  {journal} {\bibinfo
  {journal} {JHEP}\ }\textbf {\bibinfo {volume} {08}},\ \bibinfo {pages}
  {114}},\ \Eprint {https://arxiv.org/abs/1707.07679} {arXiv:1707.07679
  [hep-th]} \BibitemShut {NoStop}%
\bibitem [{\citenamefont {Feigin}\ \emph {et~al.}(2007)\citenamefont {Feigin},
  \citenamefont {Feigin},\ and\ \citenamefont {Tipunin}}]{Feigin:2007sp}%
  \BibitemOpen
  \bibfield  {author} {\bibinfo {author} {\bibfnamefont {B.}~\bibnamefont
  {Feigin}}, \bibinfo {author} {\bibfnamefont {E.}~\bibnamefont {Feigin}},\
  and\ \bibinfo {author} {\bibfnamefont {I.}~\bibnamefont {Tipunin}},\
  }\bibfield  {title} {\bibinfo {title} {{Fermionic formulas for (1,p)
  logarithmic model characters in Phi{2,1} quasiparticle realisation}},\
  }\href@noop {} {\  (\bibinfo {year} {2007})},\ \Eprint
  {https://arxiv.org/abs/0704.2464} {arXiv:0704.2464 [hep-th]} \BibitemShut
  {NoStop}%
\bibitem [{\citenamefont {Feigin}\ and\ \citenamefont
  {Tipunin}(2008)}]{feigin2008characters}%
  \BibitemOpen
  \bibfield  {author} {\bibinfo {author} {\bibfnamefont {B.~L.}\ \bibnamefont
  {Feigin}}\ and\ \bibinfo {author} {\bibfnamefont {I.~Y.}\ \bibnamefont
  {Tipunin}},\ }\href@noop {} {\bibinfo {title} {Characters of coinvariants in
  (1,p) logarithmic models}} (\bibinfo {year} {2008}),\ \Eprint
  {https://arxiv.org/abs/0805.4096} {arXiv:0805.4096 [math.QA]} \BibitemShut
  {NoStop}%
\bibitem [{\citenamefont {Buican}\ and\ \citenamefont
  {Nishinaka}(2016{\natexlab{a}})}]{Buican:2016arp}%
  \BibitemOpen
  \bibfield  {author} {\bibinfo {author} {\bibfnamefont {M.}~\bibnamefont
  {Buican}}\ and\ \bibinfo {author} {\bibfnamefont {T.}~\bibnamefont
  {Nishinaka}},\ }\bibfield  {title} {\bibinfo {title} {{Conformal Manifolds in
  Four Dimensions and Chiral Algebras}},\ }\href
  {https://doi.org/10.1088/1751-8113/49/46/465401} {\bibfield  {journal}
  {\bibinfo  {journal} {J. Phys. A}\ }\textbf {\bibinfo {volume} {49}},\
  \bibinfo {pages} {465401} (\bibinfo {year} {2016}{\natexlab{a}})},\ \Eprint
  {https://arxiv.org/abs/1603.00887} {arXiv:1603.00887 [hep-th]} \BibitemShut
  {NoStop}%
\bibitem [{\citenamefont {Cecotti}\ \emph {et~al.}(2010)\citenamefont
  {Cecotti}, \citenamefont {Neitzke},\ and\ \citenamefont
  {Vafa}}]{Cecotti:2010fi}%
  \BibitemOpen
  \bibfield  {author} {\bibinfo {author} {\bibfnamefont {S.}~\bibnamefont
  {Cecotti}}, \bibinfo {author} {\bibfnamefont {A.}~\bibnamefont {Neitzke}},\
  and\ \bibinfo {author} {\bibfnamefont {C.}~\bibnamefont {Vafa}},\ }\bibfield
  {title} {\bibinfo {title} {{R-Twisting and 4d/2d Correspondences}},\
  }\href@noop {} {\  (\bibinfo {year} {2010})},\ \Eprint
  {https://arxiv.org/abs/1006.3435} {arXiv:1006.3435 [hep-th]} \BibitemShut
  {NoStop}%
\bibitem [{Note2()}]{Note2}%
  \BibitemOpen
  \bibinfo {note} {This type of gauging realization was also observed in \cite
  {Carta:2021whq}.}\BibitemShut {Stop}%
\bibitem [{Note3()}]{Note3}%
  \BibitemOpen
  \bibinfo {note} {In this letter, we mainly focus on the local properties of
  the theories and do not care about their global forms, so we will not
  distinguish $SU(2)$ and $SO(3)$.}\BibitemShut {Stop}%
\bibitem [{\citenamefont {Cecotti}\ \emph {et~al.}(2013)\citenamefont
  {Cecotti}, \citenamefont {Del~Zotto},\ and\ \citenamefont
  {Giacomelli}}]{Cecotti:2013lda}%
  \BibitemOpen
  \bibfield  {author} {\bibinfo {author} {\bibfnamefont {S.}~\bibnamefont
  {Cecotti}}, \bibinfo {author} {\bibfnamefont {M.}~\bibnamefont {Del~Zotto}},\
  and\ \bibinfo {author} {\bibfnamefont {S.}~\bibnamefont {Giacomelli}},\
  }\bibfield  {title} {\bibinfo {title} {{More on the N=2 superconformal
  systems of type $D_p(G)$}},\ }\href {https://doi.org/10.1007/JHEP04(2013)153}
  {\bibfield  {journal} {\bibinfo  {journal} {JHEP}\ }\textbf {\bibinfo
  {volume} {04}},\ \bibinfo {pages} {153}},\ \Eprint
  {https://arxiv.org/abs/1303.3149} {arXiv:1303.3149 [hep-th]} \BibitemShut
  {NoStop}%
\bibitem [{\citenamefont {Xie}(2013)}]{Xie:2012hs}%
  \BibitemOpen
  \bibfield  {author} {\bibinfo {author} {\bibfnamefont {D.}~\bibnamefont
  {Xie}},\ }\bibfield  {title} {\bibinfo {title} {{General Argyres-Douglas
  Theory}},\ }\href {https://doi.org/10.1007/JHEP01(2013)100} {\bibfield
  {journal} {\bibinfo  {journal} {JHEP}\ }\textbf {\bibinfo {volume} {01}},\
  \bibinfo {pages} {100}},\ \Eprint {https://arxiv.org/abs/1204.2270}
  {arXiv:1204.2270 [hep-th]} \BibitemShut {NoStop}%
\bibitem [{\citenamefont {Di~Pietro}\ and\ \citenamefont
  {Komargodski}(2014)}]{DiPietro:2014bca}%
  \BibitemOpen
  \bibfield  {author} {\bibinfo {author} {\bibfnamefont {L.}~\bibnamefont
  {Di~Pietro}}\ and\ \bibinfo {author} {\bibfnamefont {Z.}~\bibnamefont
  {Komargodski}},\ }\bibfield  {title} {\bibinfo {title} {{Cardy formulae for
  SUSY theories in $d =$ 4 and $d =$ 6}},\ }\href
  {https://doi.org/10.1007/JHEP12(2014)031} {\bibfield  {journal} {\bibinfo
  {journal} {JHEP}\ }\textbf {\bibinfo {volume} {12}},\ \bibinfo {pages}
  {031}},\ \Eprint {https://arxiv.org/abs/1407.6061} {arXiv:1407.6061 [hep-th]}
  \BibitemShut {NoStop}%
\bibitem [{\citenamefont {Buican}\ and\ \citenamefont
  {Nishinaka}(2016{\natexlab{b}})}]{Buican:2015ina}%
  \BibitemOpen
  \bibfield  {author} {\bibinfo {author} {\bibfnamefont {M.}~\bibnamefont
  {Buican}}\ and\ \bibinfo {author} {\bibfnamefont {T.}~\bibnamefont
  {Nishinaka}},\ }\bibfield  {title} {\bibinfo {title} {{On the superconformal
  index of Argyres\textendash{}Douglas theories}},\ }\href
  {https://doi.org/10.1088/1751-8113/49/1/015401} {\bibfield  {journal}
  {\bibinfo  {journal} {J. Phys. A}\ }\textbf {\bibinfo {volume} {49}},\
  \bibinfo {pages} {015401} (\bibinfo {year} {2016}{\natexlab{b}})},\ \Eprint
  {https://arxiv.org/abs/1505.05884} {arXiv:1505.05884 [hep-th]} \BibitemShut
  {NoStop}%
\bibitem [{\citenamefont {Song}\ \emph {et~al.}(2017)\citenamefont {Song},
  \citenamefont {Xie},\ and\ \citenamefont {Yan}}]{Song:2017oew}%
  \BibitemOpen
  \bibfield  {author} {\bibinfo {author} {\bibfnamefont {J.}~\bibnamefont
  {Song}}, \bibinfo {author} {\bibfnamefont {D.}~\bibnamefont {Xie}},\ and\
  \bibinfo {author} {\bibfnamefont {W.}~\bibnamefont {Yan}},\ }\bibfield
  {title} {\bibinfo {title} {{Vertex operator algebras of Argyres-Douglas
  theories from M5-branes}},\ }\href {https://doi.org/10.1007/JHEP12(2017)123}
  {\bibfield  {journal} {\bibinfo  {journal} {JHEP}\ }\textbf {\bibinfo
  {volume} {12}},\ \bibinfo {pages} {123}},\ \Eprint
  {https://arxiv.org/abs/1706.01607} {arXiv:1706.01607 [hep-th]} \BibitemShut
  {NoStop}%
\bibitem [{\citenamefont {Xie}\ and\ \citenamefont {Yan}(2021)}]{Xie:2019zlb}%
  \BibitemOpen
  \bibfield  {author} {\bibinfo {author} {\bibfnamefont {D.}~\bibnamefont
  {Xie}}\ and\ \bibinfo {author} {\bibfnamefont {W.}~\bibnamefont {Yan}},\
  }\bibfield  {title} {\bibinfo {title} {{Schur sector of Argyres-Douglas
  theory and $W$-algebra}},\ }\href
  {https://doi.org/10.21468/SciPostPhys.10.3.080} {\bibfield  {journal}
  {\bibinfo  {journal} {SciPost Phys.}\ }\textbf {\bibinfo {volume} {10}},\
  \bibinfo {pages} {080} (\bibinfo {year} {2021})},\ \Eprint
  {https://arxiv.org/abs/1904.09094} {arXiv:1904.09094 [hep-th]} \BibitemShut
  {NoStop}%
\bibitem [{\citenamefont {Buican}\ and\ \citenamefont
  {Nishinaka}(2022)}]{Buican:2020moo}%
  \BibitemOpen
  \bibfield  {author} {\bibinfo {author} {\bibfnamefont {M.}~\bibnamefont
  {Buican}}\ and\ \bibinfo {author} {\bibfnamefont {T.}~\bibnamefont
  {Nishinaka}},\ }\bibfield  {title} {\bibinfo {title} {{$ \mathcal{N} $ = 4
  SYM, Argyres-Douglas theories, and an exact graded vector space
  isomorphism}},\ }\href {https://doi.org/10.1007/JHEP04(2022)028} {\bibfield
  {journal} {\bibinfo  {journal} {JHEP}\ }\textbf {\bibinfo {volume} {04}},\
  \bibinfo {pages} {028}},\ \Eprint {https://arxiv.org/abs/2012.13209}
  {arXiv:2012.13209 [hep-th]} \BibitemShut {NoStop}%
\bibitem [{\citenamefont {Jiang}(2024)}]{Jiang:2024baj}%
  \BibitemOpen
  \bibfield  {author} {\bibinfo {author} {\bibfnamefont {H.}~\bibnamefont
  {Jiang}},\ }\bibfield  {title} {\bibinfo {title} {{Modularity in
  Argyres-Douglas theories with a = c}},\ }\href
  {https://doi.org/10.1007/JHEP06(2024)131} {\bibfield  {journal} {\bibinfo
  {journal} {JHEP}\ }\textbf {\bibinfo {volume} {06}},\ \bibinfo {pages}
  {131}},\ \Eprint {https://arxiv.org/abs/2403.05323} {arXiv:2403.05323
  [hep-th]} \BibitemShut {NoStop}%
\bibitem [{\citenamefont {Arakawa}\ and\ \citenamefont
  {Moreau}(2021)}]{ArakawaMoreau2021}%
  \BibitemOpen
  \bibfield  {author} {\bibinfo {author} {\bibfnamefont {T.}~\bibnamefont
  {Arakawa}}\ and\ \bibinfo {author} {\bibfnamefont {A.}~\bibnamefont
  {Moreau}},\ }\href
  {https://www.imo.universite-paris-saclay.fr/~anne.moreau/CEMPI-arc_space-vertex_algebras.WEB_VERSION.pdf}
  {\bibinfo {title} {Arc spaces and vertex algebras}},\ \bibinfo {howpublished}
  {CEMPI Lecture Series} (\bibinfo {year} {2021}),\ \bibinfo {note} {monograph
  / Lecture Notes}\BibitemShut {NoStop}%
\bibitem [{\citenamefont {Bhargava}\ \emph {et~al.}(2024)\citenamefont
  {Bhargava}, \citenamefont {Buican},\ and\ \citenamefont
  {Jiang}}]{Bhargava:2023hsc}%
  \BibitemOpen
  \bibfield  {author} {\bibinfo {author} {\bibfnamefont {C.}~\bibnamefont
  {Bhargava}}, \bibinfo {author} {\bibfnamefont {M.}~\bibnamefont {Buican}},\
  and\ \bibinfo {author} {\bibfnamefont {H.}~\bibnamefont {Jiang}},\ }\bibfield
   {title} {\bibinfo {title} {{Exact Operator Map from Strong Coupling to Free
  Fields: Beyond Seiberg-Witten Theory}},\ }\href
  {https://doi.org/10.1103/PhysRevLett.132.031602} {\bibfield  {journal}
  {\bibinfo  {journal} {Phys. Rev. Lett.}\ }\textbf {\bibinfo {volume} {132}},\
  \bibinfo {pages} {031602} (\bibinfo {year} {2024})},\ \Eprint
  {https://arxiv.org/abs/2306.05507} {arXiv:2306.05507 [hep-th]} \BibitemShut
  {NoStop}%
\bibitem [{\citenamefont {Jiang}(2026)}]{Jiang:2026ffv}%
  \BibitemOpen
  \bibfield  {author} {\bibinfo {author} {\bibfnamefont {H.}~\bibnamefont
  {Jiang}},\ }\bibfield  {title} {\bibinfo {title} {{Macdonald index from
  vertex-operator algebras and graded unitarity}},\ }\href
  {https://doi.org/10.1103/x3rs-mf4f} {\bibfield  {journal} {\bibinfo
  {journal} {Phys. Rev. D}\ }\textbf {\bibinfo {volume} {114}},\ \bibinfo
  {pages} {025022} (\bibinfo {year} {2026})},\ \Eprint
  {https://arxiv.org/abs/2603.29829} {arXiv:2603.29829 [hep-th]} \BibitemShut
  {NoStop}%
\bibitem [{\citenamefont {Jiang}()}]{Jiang2026}%
  \BibitemOpen
  \bibfield  {author} {\bibinfo {author} {\bibfnamefont {H.}~\bibnamefont
  {Jiang}},\ }\bibfield  {title} {\bibinfo {title} {in preparation},\
  }\href@noop {} {\ }\BibitemShut {NoStop}%
\bibitem [{Note4()}]{Note4}%
  \BibitemOpen
  \bibinfo {note} {For $m=1$, this expression agrees with the proposal of
  Ref.~\cite {Yoshida:2026ugm}.}\BibitemShut {Stop}%
\bibitem [{\citenamefont {Deb}\ \emph {et~al.}(2025)\citenamefont {Deb},
  \citenamefont {Meneghelli},\ and\ \citenamefont {Rastelli}}]{Deb:2025cqr}%
  \BibitemOpen
  \bibfield  {author} {\bibinfo {author} {\bibfnamefont {A.}~\bibnamefont
  {Deb}}, \bibinfo {author} {\bibfnamefont {C.}~\bibnamefont {Meneghelli}},\
  and\ \bibinfo {author} {\bibfnamefont {L.}~\bibnamefont {Rastelli}},\
  }\bibfield  {title} {\bibinfo {title} {{The Nilpotency Index for 4d
  $\mathcal{N}=2$ SCFTs}},\ }\href@noop {} {\  (\bibinfo {year} {2025})},\
  \Eprint {https://arxiv.org/abs/2503.05975} {arXiv:2503.05975 [hep-th]}
  \BibitemShut {NoStop}%
\bibitem [{\citenamefont {Chen}\ \emph {et~al.}(2026)\citenamefont {Chen},
  \citenamefont {Deb},\ and\ \citenamefont {Rastelli}}]{Chen:2026vyz}%
  \BibitemOpen
  \bibfield  {author} {\bibinfo {author} {\bibfnamefont {S.}~\bibnamefont
  {Chen}}, \bibinfo {author} {\bibfnamefont {A.}~\bibnamefont {Deb}},\ and\
  \bibinfo {author} {\bibfnamefont {L.}~\bibnamefont {Rastelli}},\ }\bibfield
  {title} {\bibinfo {title} {{Higgsless Lagrangian SCFTs and Strongly Finite
  VOAs}},\ }\href@noop {} {\  (\bibinfo {year} {2026})},\ \Eprint
  {https://arxiv.org/abs/2607.08813} {arXiv:2607.08813 [hep-th]} \BibitemShut
  {NoStop}%
\bibitem [{\citenamefont {Yoshida}(2026)}]{Yoshida:2026ugm}%
  \BibitemOpen
  \bibfield  {author} {\bibinfo {author} {\bibfnamefont {Y.}~\bibnamefont
  {Yoshida}},\ }\bibfield  {title} {\bibinfo {title} {{SCFT/VOA correspondence
  and R-twisted reductions of $(A_2,D_{3n-2})$ Argyres--Douglas theories}},\
  }\href@noop {} {\  (\bibinfo {year} {2026})},\ \Eprint
  {https://arxiv.org/abs/2609.04099} {arXiv:2609.04099 [hep-th]} \BibitemShut
  {NoStop}%
\bibitem [{\citenamefont {Carta}\ \emph {et~al.}(2021)\citenamefont {Carta},
  \citenamefont {Giacomelli}, \citenamefont {Mekareeya},\ and\ \citenamefont
  {Mininno}}]{Carta:2021whq}%
  \BibitemOpen
  \bibfield  {author} {\bibinfo {author} {\bibfnamefont {F.}~\bibnamefont
  {Carta}}, \bibinfo {author} {\bibfnamefont {S.}~\bibnamefont {Giacomelli}},
  \bibinfo {author} {\bibfnamefont {N.}~\bibnamefont {Mekareeya}},\ and\
  \bibinfo {author} {\bibfnamefont {A.}~\bibnamefont {Mininno}},\ }\bibfield
  {title} {\bibinfo {title} {{Conformal manifolds and 3d mirrors of
  Argyres-Douglas theories}},\ }\href {https://doi.org/10.1007/JHEP08(2021)015}
  {\bibfield  {journal} {\bibinfo  {journal} {JHEP}\ }\textbf {\bibinfo
  {volume} {08}},\ \bibinfo {pages} {015}},\ \Eprint
  {https://arxiv.org/abs/2105.08064} {arXiv:2105.08064 [hep-th]} \BibitemShut
  {NoStop}%
\bibitem [{\citenamefont {Couzens}\ \emph {et~al.}(2023)\citenamefont
  {Couzens}, \citenamefont {Kang}, \citenamefont {Lawrie},\ and\ \citenamefont
  {Lee}}]{Couzens:2023kyf}%
  \BibitemOpen
  \bibfield  {author} {\bibinfo {author} {\bibfnamefont {C.}~\bibnamefont
  {Couzens}}, \bibinfo {author} {\bibfnamefont {M.~J.}\ \bibnamefont {Kang}},
  \bibinfo {author} {\bibfnamefont {C.}~\bibnamefont {Lawrie}},\ and\ \bibinfo
  {author} {\bibfnamefont {Y.}~\bibnamefont {Lee}},\ }\bibfield  {title}
  {\bibinfo {title} {{Holographic duals of Higgsed $\mathcal{D}_p^b(BCD)$}},\
  }\href@noop {} {\  (\bibinfo {year} {2023})},\ \Eprint
  {https://arxiv.org/abs/2312.12503} {arXiv:2312.12503 [hep-th]} \BibitemShut
  {NoStop}%
\bibitem [{\citenamefont {Chacaltana}\ and\ \citenamefont
  {Distler}(2013)}]{Chacaltana:2011ze}%
  \BibitemOpen
  \bibfield  {author} {\bibinfo {author} {\bibfnamefont {O.}~\bibnamefont
  {Chacaltana}}\ and\ \bibinfo {author} {\bibfnamefont {J.}~\bibnamefont
  {Distler}},\ }\bibfield  {title} {\bibinfo {title} {{Tinkertoys for the $D_N$
  series}},\ }\href {https://doi.org/10.1007/JHEP02(2013)110} {\bibfield
  {journal} {\bibinfo  {journal} {JHEP}\ }\textbf {\bibinfo {volume} {02}},\
  \bibinfo {pages} {110}},\ \Eprint {https://arxiv.org/abs/1106.5410}
  {arXiv:1106.5410 [hep-th]} \BibitemShut {NoStop}%
\end{thebibliography}%
 
\onecolumngrid
 \newpage  
 
 \begin{center} { {\Large Supplemental Materials}}\end{center}
 
 \appendix
 
 \section{central charges and levels for building blocks}\label{chargess}

We first derive the central charges and level of the right-hand building block $\fkR_m$ theory, which can be regarded as the IR theory obtained  from  the UV theory $D_{2m+1}(SO(2m+2))$   via nilpotent Higgsing. 

Let us compute the characteristic data of     UV theory $D_{2m+1}(SO(2m+2))$. The UV theory has flavor symmetry $SO(2m+2)$.  The flavor level and $c$-central charge are 
\be
 k^\text{UV} =2(2m-\frac{2m}{2m+1})=\frac{8m^2}{2m+1},
\qquad
c^\text{UV}=\frac16m(m+1)(2m+1) .
\ee
 
The Coulomb branch spectrum itself can be computed using 
\be
\cS_\text{CB}= \left\{ d - \frac{b}{p} \ell > 1 \;\middle|\; d \in \mathrm{Cas}(SO(2m+2)), \; \ell \ge 1 \right\},
\ee 
where   $
\mathrm{Cas}(SO(2m+2))=\{2,4,\cdots, 2m  ,m+1\},  b=2m,   p=2m+1$. Note that the last Pfaffian degree \(m+1\) in $\mathrm{Cas}(SO(2m+2))$ is counted separately. Then one can show that the $a$-central charge is 
\be
a^\text{UV}=\frac{1}{24} \left(8 m^3+11 m^2+3 m\right), 
\ee
by using the following formula
\be\label{TSid}
4(2a - c) = \sum_ {\Delta_j\in \cS_\text{CB}} \bigl( 2\Delta_j - 1 \bigr).
\ee
The $D_{2m+1}(SO(2m+2))$  corresponds to the full puncture, and we need to partially close the puncture in order to get the IR theory 
$\fkR_m=D_{2m+1}^{2m}(SO(2m+2),[2m-1,1^3])$.  In general, the nilpotent orbit of $SO(N)$ is labelled by the  partition $[N^{\fkm_N}\cdots 1^{\fkm_1}]$ subject to  $\sum_k k \fkm_k=N$ together with the condition that $\fkm_k$ must be even for even $k$. Our case corresponds to
$
\fkm_1=3,  \fkm_{2m-1}=1
$.
 
 The levels and central charges of the UV and IR theories are related via anomaly matching: 
\be
\begin{aligned}
24(c^{\text{IR}} - a^{\text{IR}}) &= 24(c^{\text{UV}} - a^{\text{UV}}) - H, \\
4(2a^{\text{IR}} - c^{\text{IR}}) &= 4(2a^{\text{UV}} - c^{\text{UV}}) - I_X k_G^{\text{UV}} + H_v, \\
k_i^{\text{IR}} &= I_i k ^{\text{UV}} - H_i. \\ 
\end{aligned}
\ee
The relevant quantities can be computed, e.g   using the formula  in \cite{Couzens:2023kyf}
\be 
H=m^2+m-2,\quad H_v = \frac{8m^4 - 20m^3 + 7m^2 + 11m - 6}{3}, \quad I_X=\frac{m(m-1)(2m-1)}{3}, 
\quad I_{i=1}=1,\quad H_{i=1}=4(m-1).
\ee
This then gives  the   central charges for the IR theory
\be
a^\text{IR}
=
\frac{36m^2+2m-5}{12(2m+1)},
\qquad
c^\text{IR}
=
\frac{9m^2+m-1}{3(2m+1)},
\qquad
k^\text{IR}= \frac{4(m+1)}{2m+1}.
\ee
The difference between the UV and IR data can be attributed to  the Nambu--Goldstone modes in the Higgsing process. 

In the above formula, the IR theory has flavor symmetry $\fks\fko(3)$. To translate to the $\fks\fku(2)$, we need to rescale the level by a factor of 2 and get   

\begin{equation}
a_{\mathfrak R_m}
=
\frac{36m^2+2m-5}{12(2m+1)},
\qquad
c_{\mathfrak R_m}
=
\frac{9m^2+m-1}{3(2m+1)},
 \qquad
 \qquad
k_{\fkR_m}= \frac{8(m+1)}{2m+1}.
\end{equation}

The analysis generalizes straightforwardly to $\fkR_{m,s}=  D_{(2s-1)(2m+1)}^{2(2s-1)m} \bigl( SO(2(2s-1)m+2),   [ 2(2s-1)m-1,1^3 ] \bigr)$ and we find that 
\beqn
c_{\fkR_{m,s}}
&=&
\frac{
8m^2s^3+12m^2s^2+4m^2s-6m^2
+6ms^2+5ms-9m
+s-3
}{
6(2m+1)
}, 
\\
a_{\fkR_{m,s}}
&=&
\frac{
32m^2s^3+48m^2s^2+16m^2s-24m^2
+24ms^2+19ms-39m
+5s-15
}{
24(2m+1)
} 
,
\\
k_{\fkR_{m,s}}&=& \frac{8(m+1)}{2m+1}.
\eeqn

Next, we derive the Coulomb branch spectrum of $\fkR_m$ mainly following the prescription in \cite{Chacaltana:2011ze}. Let us set $N=m+1$ and consider the $D_N$ Hitchin system with spectral curve 
\be
0=x^{2N}+\sum_{k=1}^{N-1}\phi_{2k} x^{2N-2k}+\tilde \phi_N(z)^2, \qquad \tilde \phi_N(z)^2\equiv \phi_{2N}.
\ee

For the partition   considered here $\rho=[2N-3,1,1,1]$, we need to consider the transpose and $D$-collapse
\be
[2N-3,1,1,1]\xrightarrow{\mathrm{transpose}}  [4,1^{2N-4}] \xrightarrow{D-\mathrm{collapse}} [3,1^{2N-3}],
\ee
which is just the Spaltenstein map of nilpotent orbit. Given $[3,1^{2N-3}]$, we label the boxes as  $p_i=0,1,2,2^{2N-3}$ where we label each row with only one box by 2. For $D_N$, we only need to consider even degrees. So $p_2=1, p_{2k}=2$ for $k=2, \cdots, N-1$ and $p_{2N}=2$. Since $\phi_{2N}=\tilde \phi_N^2$, we have $p_{2N}=2\tilde p_N=2$, hence $\tilde p_N=1$. So we end up with the pole structure 
\be
\{p_2, p_4, \cdots, p_{2m}; \tilde p_{m+1} \}=\{1,2^{m-1}; 1\}.
\ee
where the value of $p_i$ is   the order of pole in $\phi_{ i}$. More precisely $p_{2k}=2$ for $k=2, \cdots, m$ means  the following structure of $\phi$:
\be
\phi_{2k}(z) =\frac{u_2^{(2k)}}{z^2}+\frac{u_1^{(2k)}}{z }+\cdots,
\ee
which is characterized by two coefficients $u_1^{(2k)},u_2^{(2k)}$. Similarly, 
\be
\phi_2(z) =\frac{u_1^{(2)}}{z }+\cdots, \qquad \tilde\phi_{m+1}(z) =\frac{  u_1^{(m+1)}}{z }+\cdots.
 \ee
 
 These coefficients $u_j^{(d)}$  just  correspond to the Coulomb branch operators. We can compute the dimension 
 \be
 \Delta(u_j^{(d)})=d[x]+j[z]= \frac{d+2mj}{2m+1}=d- \frac{2m}{2m+1}(d-j)=\Delta_{d, j},
 \ee
 where we used $[x]=\frac{1}{2m+1},[z]=\frac{2m}{2m+1}$,   
and  introduced the notation
\be
 \Delta_{d,j}\equiv d - \frac{2m}{2m+1} (d-j).
\ee

We thus find the Coulomb branch spectrum is given by 
\be
\mathcal S_{\mathrm{CB}}(\mathfrak R_m)
=\Big\{
\Delta_{2,1},\Delta_{m+1,1}
\Big\}
\bigcup
\Big\{
\Delta_{2k, 1},\Delta_{2k, 2}\Big| k=2, \cdots, m 
\Big\},
\ee
%
which can be explicitly written as
\begin{equation}
\begin{aligned}
\mathcal S_{\mathrm{CB}}(\mathfrak R_m)
={}&
\left\{
\frac{2j}{2m+1}
\ \middle|
j=m+1,\ldots,2m
\right\}
\cup
\left\{
\frac{2j}{2m+1}
\ \middle|
j=2m+2,\ldots,3m
\right\} 
\cup
\left\{
\frac{3m+1}{2m+1}
\right\}.
\end{aligned}
\end{equation}
It is easy to check that the spectrum and central charges are   consistent with  \eqref{TSid}.

Now we switch to the left-hand theory $\fkL_m=D_{2m+1}(SO(3))=D_{2m+1}(SU(2))=(A_1,D_{2m+1})$. The $D_{2m+1}(SU(2))$ presentation allows us to perform the same type of computation.  In particular the Coulomb branch spectrum is
\be 
\cS_\text{CB}(\fkL _m)= \left\{ d - \frac{2}{2m+1} \ell > 1 \;\middle|\; d =2, \; \ell \ge 1 \right\}
=
\left\{
\frac{2j}{2m+1}
\ \middle|
j=m+1,\ldots,2m
\right\},
\ee 
and the remaining data are
\begin{equation}
a_{\mathfrak L_m}
=
\frac{m(8m+3)}{8(2m+1)},
\qquad
c_{\mathfrak L_m}
=
\frac{m}{2},
 \qquad
k_{\mathfrak L_m}^{4d}
=
\frac{8m}{2m+1}.
\end{equation} 
The associated 
VOA of $\cL_m$ is just the affine Kac-Moody algebra $\widehat{\fks\fku(2)}_{-\frac{4m}{2m+1}}$, whose level and central charge of the Sugawara stress tensor indeed reproduce the results above, after translating  from 2d to 4d. 

 Finally, the data of the  free vector multiplet in the adjoint representation of $\fks\fku(2)$ are 
\be
\mathcal S_{\mathrm{CB}}(\mathrm{vec})=\{2\}, \qquad a_{\mathrm{vec}}=\frac58,
\qquad
c_{\mathrm{vec}}=\frac12.
\ee

\section{Proof of the identity between index and character }\label{ISischiapp}

We now prove that equality \eqref{ISischi} between Schur index \eqref{IschurTm} and character
\eqref{superch}, namely:
\beqn
\cI^S_{\cT_m}(q)&\equiv& \label{LHSeq}
\frac12 
\PE\Big[ 
\frac{(q^2-q^{2m})
}{
(1-q)(1-q^{2m+1})}
\Big]
\oint\frac{dz}{2\pi i z}(1-z^2)(1-z^{-2})
\times\PE\Big[ 
\frac{ 
q^m(1+q-2 q^{m+1})
\chi_{\mathbf 3}(z)
}{
 (1-q^{2m+1})
}
\Big]
\\&=&  \label{RHSeq}
\frac{1}{(q;q)_\infty}
\sum_{r=1}^{\infty}
(-1)^{r-1}r\,q^{\frac{(r-1)((2m+1)r+2m)}{2}}
(1-q^r) \equiv \chi_{\cA(4m+2)}(q).
\eeqn

\subsection{Theta-function and index}

For later convenience, we set
\begin{equation}
Q:=q^{2m+1},
\qquad
a:=q^m,
\qquad
b:=q^{m+1}=\frac{Q}{a},
\qquad
x:=z^2.
\end{equation}

Starting with the  $q$-Pochhammer symbol  $(x;Q)_\infty=\prod_{n=0}^\infty (1-xQ^n)$, let us   define the following theta function
\begin{equation}  \label{thetadef}
\theta(x;Q)
:=
(x;Q)_\infty
(Q/x;Q)_\infty
(Q;Q)_\infty,\qquad 
\end{equation}
where    the product of the first two terms on the RHS is the $q$-theta function, while the last factor is included  for convenience. 
It satisfies
\be\label{thetapd}
\theta(x;Q)=\theta(\frac{Q}{x}; Q) =-x\; \theta(\frac{1}{x};Q)=-x \;\theta( {Q}{x};Q),
\ee
and its derivative at $x=1$ is
\begin{align} \label{thed}
\theta'(1;Q)
&=
\p_x ((1-x)(Qx;Q)_\infty(Q/x;Q)_\infty(Q;Q)_\infty)|_{x=1}
=
-(Q;Q)_\infty^3.
\end{align}

For a Weyl-invariant Laurent series \(F(x)=F(x^{-1})\), the normalized
\(SU(2)\) Haar integration is
\begin{equation}\label{Hmea}
\int d\mu_{SU(2)}(z)\,F(z^2)\equiv \frac12 \oint\frac{dz}{2\pi i z}(1-z^2)(1-z^{-2})F(z^2)
=
\CT_x\left[(1-x)F(x)\right]. 
\end{equation}
where the constant term is defined as:
\be
\CT_x G(x) =\oint\frac{dx }{2\pi i x}G(x).
\ee
 
We first note that
\beqn \label{Pex}
\PE\Big[\frac{x}{1-Q}\Big] &=& \PE\Big[\sum_{n=0}^\infty x Q^n\Big]=\prod_{n=0}^\infty  \PE[xQ^n]=\prod_{n=0}^\infty  \frac{1}{1-xQ^n}
=\frac{1}{(x;Q)_\infty} ,
\\  \label{Pemx}
 \PE\Big[\frac{-x}{1-Q}\Big]&=&  \PE\Big[\sum_{n=0}^\infty -x Q^n\Big]=\prod_{n=0}^\infty  \PE[-xQ^n]=\prod_{n=0}^\infty ({1-xQ^n})
= {(x;Q)_\infty}  .
\eeqn

Then it is easy to show that
\be
\PE\Big[ 
\frac{(q^2-q^{2m})
}{
(1-q)(1-q^{2m+1})}
\Big]
=
\PE\Big[ 
\frac {q^2(1- q^{2m-2}) 
}{
(1-q)(1-q^{2m+1})}
\Big]
=
\PE\Big[ 
\frac {q^2\sum_{n=0}^{2m-3 } q^n
}{
 (1-q^{2m+1})}
\Big]
=
\prod_{j=2}^{2m-1}
\frac{1}{(q^j;Q)_\infty} \equiv S(q) ,
\label{PES}
\ee
where we introduce the notation $S(q)$ for later convenience.

Furthermore, we have 
\be
 \frac{-2q^{2m+1}}{1-q^{2m+1}} +\frac{q^m(1+q)}{1-q^{2m+1}}  
=\frac{a+b-2Q}{1-Q},
\ee
which implies 
\be
\Big( \frac{-2q^{2m+1}}{1-q^{2m+1}} +\frac{q^m(1+q)}{1-q^{2m+1}} \Big)\chi_{\bm 3}(z)
=\frac{a+b-2Q}{1-Q}+\frac{a+b-2Q}{1-Q}x+\frac{a+b-2Q}{1-Q}\frac{1}{x} .
\ee

Using \eqref{Pex} and \eqref{Pemx}, we get
\beqn 
&&\PE\Big[ \Big( \frac{-2q^{2m+1}}{1-q^{2m+1}} +\frac{q^m(1+q)}{1-q^{2m+1}}  \Big)\chi_{\bm 3}(z)
\Big]
= 
\frac{(Q;Q)_\infty^2}
{(a;Q)_\infty(b;Q)_\infty}
\frac{
(Qx;Q)_\infty^2(Q/x;Q)_\infty^2
}{
(ax;Q)_\infty(bx;Q)_\infty
(a/x;Q)_\infty(b/x;Q)_\infty}.
\label{PEins}
\eeqn

Combining \eqref{PEins}, \eqref{PES} and \eqref{Hmea}, the index  \eqref{LHSeq} then can be written as
\begin{align}
\cI^S_{\cT_m} (q)
={}&
S (q)
\frac{(Q;Q)_\infty^2}
{(a;Q)_\infty(b;Q)_\infty}
\times\CT_x
\left[
(1-x)
\frac{
(Qx;Q)_\infty^2(Q/x;Q)_\infty^2
}{
(ax;Q)_\infty(bx;Q)_\infty
(a/x;Q)_\infty(b/x;Q)_\infty
}
\right].
\label{eqindx}
\end{align}

Using
\[
(x;Q)_\infty=(1-x)(Qx;Q)_\infty,
\]
and \(ab=Q \), the constant-term integrand reduces to
\beqn
(1-x)
\frac{
(Qx;Q)_\infty^2(Q/x;Q)_\infty^2
}{
(ax;Q)_\infty(bx;Q)_\infty
(a/x;Q)_\infty(b/x;Q)_\infty
}
&=&
\frac{
( x;Q)_\infty^2(Q/x;Q)_\infty^2
}{
(1-x) (ax;Q)_\infty(bx;Q)_\infty
(a/x;Q)_\infty(b/x;Q)_\infty
}
\\&=&
\frac{
\theta(x;Q)^2
}{
(1-x)\theta(ax;Q)\theta(a/x;Q)
}.
\eeqn
This further simplifies \eqref{eqindx} and yields
\begin{align}
\cI^S_{\cT_m}(q)
={}&
S (q)
\frac{(Q;Q)_\infty^2}
{(a;Q)_\infty(b;Q)_\infty}
\times\CT_x
\left[
\frac{
\theta(x;Q)^2}{(1-x)\;\theta(ax;Q)\theta(a/x;Q)}
\right].
\label{eqindx2}
\end{align}

\subsection{The constant-term lemma}
To proceed, we need the following lemma

\begin{equation} 
\CT_x
\frac{
\theta(x;Q)^2
}{
(1-x)\theta(ax;Q)\theta(a/x;Q)
}
=
\frac{
\theta(a;Q)\,\partial_a\theta(a;Q)
}{
\theta(a^2;Q)\theta'(1;Q)
},\qquad \qquad 0<|Q| <|a|<1. 
\label{eqlemma}
\end{equation}

It can be proved as follows. We first define
\begin{equation}
H_a(x)
:=
\frac{
\theta(x;Q)^2
}{
\theta(ax;Q)\theta(a/x;Q)
},
\end{equation}
 which satisfies 
\begin{equation}
H_a(Qx)=\frac{ \theta(Qx;Q)^2 }{ \theta(aQx;Q)\theta(a/Q/x;Q)}
=\frac{\frac{1}{x^2} \theta( x;Q)^2 }{ \frac{-a/Q/x}{-a x } \theta(a x;Q)\theta(a /x;Q)}
=QH_a(x).
\label{eq:H-quasi}
\end{equation}
where we used the identities   in \eqref{thetapd}.

Choose a radius \(\rho\) such that
$
\max{|a|,|b|}<\rho<1 
$.
Since
\be
\frac{1}{1-x}=\sum_{j=0}^{\infty}x^j,
\qquad
|x|=\rho,
\ee
we can write
\begin{equation}\label{CaQ}
C(a,Q)
:=
\CT_x\frac{H_a(x)}{1-x}
=\CT_x \sum_{j=0}^\infty  H_a(x) x^j
=
\sum_{j=0}^{\infty}M_j,
\end{equation}
where
\begin{equation}
M_j
=
\frac{1}{2\pi i}
\oint_{|x|=\rho}
H_a(x)x^{j-1}\,dx.
\end{equation}
We can now shrink the contour from \(|x|=\rho\) to
\(|x|=|Q|\rho\).  In this process, the integral is changed: 
\beqn
\frac{1}{2\pi i}
 (\oint_{|x|=\rho}-\oint_{|x|=Q\rho})
H_a(x)x^{j-1}\,dx
&=&
\frac{1}{2\pi i}
  \oint_{|x|=\rho}  (H_a(x)x^{j-1}\,dx
-Q^j H_a(Qx)x^{j-1}\,dx)
\\&=&
\frac{1}{2\pi i}
  \oint_{|x|=\rho}  (H_a(x)x^{j-1}\,dx
-Q^{j+1} H_a( x)x^{j-1}\,dx)
\\&=&
(1-Q^{j+1} )\frac{1}{2\pi i}
  \oint_{|x|=\rho}   H_a(x)x^{j-1}\,dx 
  =(1-Q^{j+1})M_j.
    \eeqn
Alternatively, the change of the integral can be computed from the residue theorem. Indeed, during the shrink of the contour, the integrand crosses the poles at   \(x=a\) and \(x=b=Q/a\). Note that $\cdots<|bQ|   <|a Q|<\rho |Q|<|Q|<|b|<|a|<\rho <1$.
The two residues are given by
\begin{align}
 {\operatorname{Res }_{x=a}}\;
H_a(x)x^{j-1} 
&=
\frac{x^{j-1} \theta(x;Q)^2}{\theta(ax;Q)\p_x \theta(a/x;Q) }\Big|_{x=a}
=
-
\frac{
a^j\theta(a;Q)^2
}{
\theta(a^2;Q)\theta'(1;Q)
},
\\
 {\operatorname{Res }_{x=b}}\;
H_a(x)x^{j-1} 
&=
\frac{x^{j-1} \theta(x;Q)^2}{\p_x\theta(ax;Q)  \theta(a/x;Q) }\Big|_{x=b}
=
\frac{x^{j-1} \theta(Q/x;Q)^2}{\p_x\theta(b/x;Q) \cdot \frac{-a}{x} \theta(Qa/x;Q) }\Big|_{x=b}
=
\frac{b^{j+1}}{a}
\frac{
\theta(a;Q)^2
}{
\theta(a^2;Q)\theta'(1;Q)
}.
\end{align}

Together, we have the equality 
\begin{equation}
(1-Q^{j+1})M_j
=
 {\operatorname{Res }_{x=a}}\;
H_a(x)x^{j-1} 
+
 {\operatorname{Res }_{x=b}}\;
H_a(x)x^{j-1} ,
\end{equation}
which implies
\begin{equation}
M_j
=
\frac{
\theta(a;Q)^2
}{
\theta(a^2;Q)\theta'(1;Q)
}
\frac{
b^{j+1}/a-a^j
}{
1-Q^{j+1}
}.
\end{equation}

Summing over \(j\geq 0\)  in \eqref{CaQ} gives
\begin{equation}\label{caqsimp}
C(a,Q)
=
\frac{
\theta(a;Q)^2
}{
a\,\theta(a^2;Q)\theta'(1;Q)
}
\sum_{n=1}^{\infty}
\frac{b^n-a^n}{1-Q^n}
.
\end{equation}

 To further simplify, we need to use the following identity: 
 \begin{equation}
a\frac{\partial_a\theta(a;Q)}{\theta(a;Q)}
=
\sum_{n=1}^{\infty}
\frac{b^n-a^n}{1-Q^n},
\end{equation}
which we prove below. 

Taking the logarithm of \eqref{thetadef} gives
\be
\log\theta(a;Q)
=
\log(Q;Q)_\infty
+\sum_{k=0}^\infty\log(1-aQ^k)
+\sum_{k=0}^\infty
\log\left(1-\frac{Q^{k+1}}{a}\right).
\ee
Further  differentiating with respect to \(a\) with $Q$ fixed yields
\be
a\partial_a\log\theta(a;Q)
=
-\sum_{k=0}^\infty
\frac{aQ^k}{1-aQ^k}
+
\sum_{k=0}^\infty
\frac{Q^{k+1}/a}{1-Q^{k+1}/a}
=
-\sum_{k=0}^\infty
\frac{aQ^k}{1-aQ^k}
+
\sum_{k=0}^\infty
\frac{bQ^{k } }{1-bQ^{k } }
.
\ee
Using
$
\frac{x}{1-x}=\sum_{n=1}^\infty x^n,
$
the two terms become 
\be
\sum_{k=0}^\infty
\frac{aQ^k}{1-aQ^k}
=
\sum_{k=0}^\infty\sum_{n=1}^\infty
a^nQ^{kn}
=
\sum_{n=1}^\infty
\frac{a^n}{1-Q^n},
\qquad\qquad
\sum_{k=0}^\infty
\frac{bQ^k}{1-bQ^k}
=
\sum_{k=0}^\infty\sum_{n=1}^\infty
b^nQ^{kn}
=
\sum_{n=1}^\infty
\frac{b^n}{1-Q^n}.
\ee
Therefore
\[
a\frac{\partial_a\theta(a;Q)}{\theta(a;Q)}
=
a\partial_a\log\theta(a;Q)
=
\sum_{n=1}^\infty
\frac{b^n-a^n}{1-Q^n}.
\]
 Substituting this into \eqref{caqsimp} gives
 \be
C(a,Q)
=
\frac{
\theta(a;Q)^2
}{
a\,\theta(a^2;Q)\theta'(1;Q)
}
\sum_{n=1}^{\infty}
\frac{b^n-a^n}{1-Q^n}
=\frac{
\theta(a;Q)\,\partial_a\theta(a;Q)
}{
\theta(a^2;Q)\theta'(1;Q)
}.
\ee

This proves the lemma \eqref{eqlemma}.
 
\subsection{Simplification of index}

Applying Lemma~\eqref{eqlemma} to
\eqref{eqindx2} and using 
 \eqref{thetadef} \eqref{thed}
gives
\begin{equation}
\cI^S_{\cT_m} (q)
=
 S (q) \frac{
(Q;Q)_\infty^3\partial_a\theta(a;Q)
}{\theta(Q/a^2;Q) \theta'(1;Q)}
=
- S (q) \frac{
\partial_a\theta(a;Q)
}{\theta(Q/a^2;Q)}.
\end{equation}
Since
$
\frac{Q}{a^2}=q,
$
this becomes
\begin{equation}
\cI^S_{\cT_m} (q)
=
-
S (q)
\frac{
\partial_a\theta(a;Q)
}{
\theta(q;Q)
}.
\end{equation}

Now using the definition in \eqref{thetadef}, it is easy to show
 \begin{equation}
\theta(q;Q)/S(q)=\theta(q;Q)\prod_{j=2}^{2m-1}(q^j;Q)_\infty 
=
 \prod_{j=1}^{2m-1}(q^j;Q)_\infty \times (q^{2m};Q)_\infty
(Q;Q)_\infty
=
(q;q)_\infty.
\end{equation}
Therefore
\begin{equation} 
\cI^S_{\cT_m} (q)
=
-\frac{
\left.
\partial_a\theta(a;Q)
\right|_{
a=q^m,
Q=q^{2m+1}
}
}{
(q;q)_\infty
}. 
\label{eq:index-theta-derivative}
\end{equation}


Next we need to use the  Jacobi triple-product identity,  which states that 
 \be
\sum_{s=-\infty}^{\infty} Q^{\frac{s(s+1)}{2}} y^s = (Q; Q)_\infty \left(-\frac{1}{y}; Q\right)_\infty (-yQ; Q)_\infty.
\ee
Setting $y=-a/Q$ gives 
\begin{equation}
\theta(a;Q)
=
\sum_{s\in\mathbb Z}
(-1)^sQ^{s(s-1)/2}a^s.
\end{equation}
Consequently,
\begin{equation}
-\partial_a\theta(a;Q)
=
\sum_{s\in\mathbb Z}
(-1)^{s-1}s\,
Q^{s(s-1)/2}a^{s-1}.
\end{equation}
At
$
Q=q^{2m+1},
a=q^m,
$
one obtains
\begin{equation}
-\partial_a\theta(a;Q)
=
\sum_{s\in\mathbb Z}
(-1)^{s-1}s\,q^{\frac{(s-1)((2m+1)s+2m)}{2}}.
\end{equation}

Pairing the terms \(s=r\) and \(s=-r\), \(r\geq 1\), yields
\begin{equation}
-\partial_a\theta(a;Q)
=
\sum_{r=1}^{\infty}
(-1)^{r-1}r\,q^{\frac{(r-1)((2m+1)r+2m)}{2}}
(1-q^r).
\end{equation}
This implies that the index in \eqref{eq:index-theta-derivative} can be written as
\be
\cI^S_{\cT_m}(q)= \frac{1}{(q;q)_\infty}
\sum_{r=1}^{\infty}
(-1)^{r-1}r\,q^{\frac{(r-1)((2m+1)r+2m)}{2}}
(1-q^r),
\ee
which is exactly the character in \eqref{superch}. This completes the  proof of the equality   \eqref{ISischi}.

 \section{Characters and modular   matrices}\label{mST}
 In this appendix, we present more details about the modular properties of the index of the $\cT_2$ theory for $m=2$. 

\subsection{MLDE}
Let us first introduce some mathematical notation. 
The Serre derivatives and $k$-th order modular differential operators  are given by 
\be
\partial_{(k)} f(q) = (q \partial_q + k \mathbb{E}_2(\tau)) f(q).
\qquad
D_q^{(k)} = \partial_{(2k-2)} \circ \cdots \circ \partial_{(2)} \circ \partial_{(0)},
\ee

 The  Eisenstein series are defined by
\[
\mathbb{E}_{2k}(\tau) = -\frac{B_{2k}}{(2k)!} + \frac{2}{(2k-1)!} \sum_{n=1}^{\infty} \frac{n^{2k-1}q^n}{1-q^n}, \qquad q \equiv e^{2\pi i\tau}.
\] 
where \( B_{2k} \) is the \( 2k \)-th Bernoulli number. 

The MLDE for $\cT_2$ is found to be
\be
\cD_q  \cZ_{\cT_2} =0,
\ee
where
 \beqn
\cD_q &\equiv& 
 D_q^{(8)} - \frac{3948}{5} \bE_4 D_q^{(6)} - \frac{44856}{5} \bE_6 D_q^{(5)} 
  + \frac{844518}{25} \bE_4^2 D_q^{(4)} 
    + \frac{10480848}{25} \bE_4 \bE_6 D_q^{(3)}     
       \nonumber   \\&&  \qquad
    + \frac{(333085204 \bE_4^3 - 315932400 \bE_6^2)}{125} D_q^{(2)} 
     +   \frac{52466568}{125} \bE_4^2 \bE_6 D_q^{(1)}
        + \frac{1}{125} (-377011635 \bE_4^4 + 1537331488 \bE_4 \bE_6^2).
        \qquad      \qquad
\eeqn

\subsection{Characters}
By making the ansatz, $\chi_b=q^b(1+\cdots)$, we get the indicial equation from the leading order of $\cD_q \chi_b=0$:
\be
(12 b-7)^2 (60 b-119) (60 b-71) (60 b-11)^2 (60 b+1)^2=0,
\ee
whose solutions are
\be
b=-\frac{1}{60},-\frac{1}{60},\frac{11}{60},\frac{11}{60},\frac{7}{12},\frac{7}{12},\frac{71}{60},\frac{119}{60}.
\ee
Correspondingly, the conformal weights   $h=b+c/24$ with $c =-\frac{238}{5}$ are
\be
h=-2,-2,-\frac{9}{5},-\frac{9}{5},-\frac{7}{5},-\frac{7}{5},-\frac{4}{5},0.
\ee

Next, we present the solutions to MLDE. Let us first define
\beqn\label{thetafcn}
\theta_{r,10}(q)
=
\sum_{n\in\mathbb Z}
q^{(20n+r)^2/40},
\qquad
\widetilde\theta_{r,10}(q)
=
\sum_{n\in\mathbb Z}
\left(n+\frac{r}{20}\right)
q^{(20n+r)^2/40}.
\eeqn

Then the five non-logarithmic solutions are
\[
\begin{aligned}
\chi_0
&=
\frac{
\frac1{10}\bigl(\theta_{9,10}-\theta_{1,10}\bigr)
+
2\bigl(\widetilde\theta_{9,10}+\widetilde\theta_{1,10}\bigr)
}{\eta}
=
q^{119/60}
\left(
1+q^2+q^3+2q^4+2q^5+4q^6+2q^7
+5q^8+6q^9+8q^{10}+\cdots
\right),
\\[4mm]
\chi_{-\frac45}
&=
\frac{
\frac3{10}\bigl(\theta_{7,10}-\theta_{3,10}\bigr)
+
2\bigl(\widetilde\theta_{7,10}+\widetilde\theta_{3,10}\bigr)
}{\eta}
=
q^{71/60}
\left(
1+q+2q^2+2q^3+4q^4+5q^5+6q^6
+8q^7+\cdots
\right),
\\[4mm]
\chi_{-\frac75}
&=
\frac{4\widetilde\theta_{5,10}}{\eta}
 =
q^{7/12}
\left(
1+q+2q^2+3q^3+5q^4+4q^5+8q^6
+9q^7+\cdots
\right),
\\[4mm]
\chi_{-\frac95}
&=
\frac{
-\frac7{10}\bigl(\theta_{7,10}-\theta_{3,10}\bigr)
+
2\bigl(\widetilde\theta_{7,10}+\widetilde\theta_{3,10}\bigr)
}{\eta}
 =
q^{11/60}
\left(
1+q+2q^2+3q^3+3q^4+5q^5+7q^6
+8q^7+\cdots
\right),
\\[4mm]
\chi_{-2}
&=
\frac{
-\frac9{10}\bigl(\theta_{9,10}-\theta_{1,10}\bigr)
+
2\bigl(\widetilde\theta_{9,10}+\widetilde\theta_{1,10}\bigr)
}{\eta}
 =
q^{-1/60}
\left(
1+q+2q^2+q^3+3q^4+3q^5+5q^6
+5q^7+\cdots
\right).
\end{aligned}
\]

We choose the logarithmic solutions such that the series multiplying
$\log q$ have primitive integral coefficients:
\[
\begin{aligned}
\widehat\chi_{-\frac75}
&=
\log q\,
\frac{4\widetilde\theta_{5,10}}{\eta}
 =
(\log q)\,
q^{7/12}
\left(
1+q+2q^2+3q^3+5q^4+4q^5+8q^6
+9q^7+\cdots
\right),
\\[4mm]
\widehat\chi_{-\frac95}
&=
\log q\,
\frac{
20\bigl(
\widetilde\theta_{7,10}
+
\widetilde\theta_{3,10}
\bigr)
}{\eta}
 =
(\log q)\,
q^{11/60}
\left(
3+10q+13q^2+23q^3+23q^4+43q^5
+56q^6+66q^7+\cdots
\right),
\\[4mm]
\widehat\chi_{-2}
&=
\log q\,
\frac{
20\bigl(
\widetilde\theta_{9,10}
+
\widetilde\theta_{1,10}
\bigr)
}{\eta}
 =
(\log q)\,
q^{-1/60}
\left(
1+q+11q^2+q^3+12q^4+12q^5
+23q^6+23q^7+\cdots
\right).
\end{aligned}
\]

 \subsection{Modular transformations}

We choose the ordered basis
\be
\boldsymbol\chi
=
\left(
\chi_0,\,
\chi_{-\frac45},\,
\chi_{-\frac75},\,
\widehat\chi_{-\frac75},\,
\chi_{-\frac95},\,
\widehat\chi_{-\frac95},\,
\chi_{-2},\,
\widehat\chi_{-2}
\right)^{T}
\ee
 
The modular transformations are
\[
\boldsymbol\chi(-1/\tau)
=
S\,\boldsymbol\chi(\tau),
\qquad
\boldsymbol\chi(\tau+1)
=
T\,\boldsymbol\chi(\tau).
\]

The explicit form of $S,T$ can be deduced from the modular transformations of the theta function in \eqref{thetafcn}. First, we note that \be\label{theta10}
\theta_{r+20,10}=\theta_{r,10},
\qquad
\theta_{-r,10}=\theta_{r,10},
\qquad
\widetilde\theta_{r+20,10}=\widetilde\theta_{r,10},
\qquad
\widetilde\theta_{-r,10}=-\widetilde\theta_{r,10},
\ee 
which implies that $\widetilde\theta_{0,10}=\widetilde\theta_{10,10}=0$.
 
Under $T:\tau\mapsto\tau+1$,
\be
\theta_{r,10}(\tau+1)
=
e^{\pi i r^2/20}\,
\theta_{r,10}(\tau)
\qquad
\widetilde\theta_{r,10}(\tau+1)
=
e^{\pi i r^2/20}\,
\widetilde\theta_{r,10}(\tau)
\ee

Under $S:\tau\mapsto-1/\tau$,
\beqn 
\theta_{r,10}\!\left(-\frac1\tau\right)
=
\sqrt{\frac{-i\tau}{20}}
\sum_{s=0}^{19}
e^{-\pi i rs/10}\,
\theta_{s,10}(\tau),
\qquad\qquad\quad
\widetilde\theta_{r,10}\!\left(-\frac1\tau\right)
=
\tau\sqrt{\frac{-i\tau}{20}}
\sum_{s=0}^{19}
e^{-\pi i rs/10}\,
\widetilde\theta_{s,10}(\tau)
\eeqn


Since
\[
\eta\!\left(-\frac1\tau\right)
=
\sqrt{-i\tau}\,\eta(\tau),
\qquad
\eta(\tau+1)
=
e^{\pi i/12}\eta(\tau),
\] 
we immediately get
\be
\frac{\theta_{r,10}}{\eta}
\Big(-\frac{1}{\tau} \Big) 
=
\frac1{\sqrt{20}}
\sum_{s=0}^{19}
e^{-\pi i rs/10}
\frac{\theta_{s,10}}{\eta}\Big( {\tau} \Big) ,
 \qquad\qquad
\frac{\widetilde\theta_{r,10}}{\eta}
\Big(-\frac{1}{\tau} \Big) 
=
\frac{\tau}{\sqrt{20}}
\sum_{s=0}^{19}
e^{-\pi i rs/10}
\frac{\widetilde\theta_{s,10}}{\eta} \Big( {\tau} \Big) 
\ee 

With these identities, we find that the modular $S$ matrix is given by
 
\begingroup
\renewcommand{\arraystretch}{1.8}
\setlength{\arraycolsep}{9pt}
\[
{\scriptsize
S=
\begin{pmatrix}
\dfrac1{10}\sqrt{\dfrac{5+\sqrt5}{10}}
&
\dfrac1{10}\sqrt{\dfrac{5-\sqrt5}{10}}
&
0
&
-\dfrac{1}{2\pi\sqrt5}
&
-\dfrac1{10}\sqrt{\dfrac{5-\sqrt5}{10}}
&
-\dfrac{5+\sqrt5}{200\pi}
&
-\dfrac1{10}\sqrt{\dfrac{5+\sqrt5}{10}}
&
-\dfrac{5-\sqrt5}{200\pi}
\\[2mm]
\dfrac3{10}\sqrt{\dfrac{5-\sqrt5}{10}}
&
-\dfrac3{10}\sqrt{\dfrac{5+\sqrt5}{10}}
&
0
&
\dfrac{1}{2\pi\sqrt5}
&
\dfrac3{10}\sqrt{\dfrac{5+\sqrt5}{10}}
&
-\dfrac{5-\sqrt5}{200\pi}
&
-\dfrac3{10}\sqrt{\dfrac{5-\sqrt5}{10}}
&
-\dfrac{5+\sqrt5}{200\pi}
\\[2mm]
0
&
0
&
0
&
-\dfrac{1}{2\pi\sqrt5}
&
0
&
\dfrac{1}{10\pi\sqrt5}
&
0
&
-\dfrac{1}{10\pi\sqrt5}
\\[2mm]
-\dfrac{18\pi}{5\sqrt5}
&
\dfrac{14\pi}{5\sqrt5}
&
-\dfrac{2\pi}{\sqrt5}
&
0
&
\dfrac{6\pi}{5\sqrt5}
&
0
&
-\dfrac{2\pi}{5\sqrt5}
&
0
\\[2mm]
-\dfrac7{10}\sqrt{\dfrac{5-\sqrt5}{10}}
&
\dfrac7{10}\sqrt{\dfrac{5+\sqrt5}{10}}
&
0
&
\dfrac{1}{2\pi\sqrt5}
&
-\dfrac7{10}\sqrt{\dfrac{5+\sqrt5}{10}}
&
-\dfrac{5-\sqrt5}{200\pi}
&
\dfrac7{10}\sqrt{\dfrac{5-\sqrt5}{10}}
&
-\dfrac{5+\sqrt5}{200\pi}
\\[2mm]
-\dfrac{9\pi(5+\sqrt5)}{5}
&
-\dfrac{7\pi(5-\sqrt5)}{5}
&
4\pi\sqrt5
&
0
&
-\dfrac{3\pi(5-\sqrt5)}{5}
&
0
&
-\dfrac{\pi(5+\sqrt5)}{5}
&
0
\\[2mm]
-\dfrac9{10}\sqrt{\dfrac{5+\sqrt5}{10}}
&
-\dfrac9{10}\sqrt{\dfrac{5-\sqrt5}{10}}
&
0
&
-\dfrac{1}{2\pi\sqrt5}
&
\dfrac9{10}\sqrt{\dfrac{5-\sqrt5}{10}}
&
-\dfrac{5+\sqrt5}{200\pi}
&
\dfrac9{10}\sqrt{\dfrac{5+\sqrt5}{10}}
&
-\dfrac{5-\sqrt5}{200\pi}
\\[2mm]
-\dfrac{9\pi(5-\sqrt5)}{5}
&
-\dfrac{7\pi(5+\sqrt5)}{5}
&
-4\pi\sqrt5
&
0
&
-\dfrac{3\pi(5+\sqrt5)}{5}
&
0
&
-\dfrac{\pi(5-\sqrt5)}{5}
&
0
\end{pmatrix}
}
\]
\endgroup
%

From \eqref{chihT}, one has  
\be
\chi_0(-1/ \tau)=-\frac{i\,(5 - \sqrt{5})}{100 }\tau \, e^{-\frac{\pi i\tau}{30 }}+\cdots=-\frac{ (5 - \sqrt{5})}{200 \pi}\log q \, q^{-\frac{1}{60 }}+\cdots
=-\frac{ (5 - \sqrt{5})}{200 \pi}\widehat\chi_{-2}+\cdots
\ee
 in the limit $q\to 0$. The coefficient $-\frac{ (5 - \sqrt{5})}{200 \pi}$ indeed agrees with the relevant entry in the $S$ matrix above.

Similarly, the $T$ matrix is given by
\begingroup
\renewcommand{\arraystretch}{1.8}
\setlength{\arraycolsep}{10pt}
\[
{\small
T=
\begin{pmatrix}
e^{-\pi i/30}
&0&0&0&0&0&0&0
\\[1mm]
0
&e^{11\pi i/30}
&0&0&0&0&0&0
\\[1mm]
0&0
&e^{7\pi i/6}
&0&0&0&0&0
\\[1mm]
0&0
&2\pi i\,e^{7\pi i/6}
&e^{7\pi i/6}
&0&0&0&0
\\[1mm]
0&0&0&0
&e^{11\pi i/30}
&0&0&0
\\[1mm]
0
&14\pi i\,e^{11\pi i/30}
&0&0
&6\pi i\,e^{11\pi i/30}
&e^{11\pi i/30}
&0&0
\\[1mm]
0&0&0&0&0&0
&e^{-\pi i/30}
&0
\\[1mm]
18\pi i\,e^{-\pi i/30}
&0&0&0&0&0
&2\pi i\,e^{-\pi i/30}
&e^{-\pi i/30}
\end{pmatrix}
}
\]
\endgroup
Note that  the diagonal phases are $e^{2\pi i(h-c/24)} $, while the non-diagonal terms  follow from
$
\log q
\longmapsto
\log q+2\pi i
$
under $\tau\mapsto\tau+1$.  

With this normalization, one can check that
\be
S^2=\mathbf 1,
\qquad
(ST)^3=\mathbf 1.
\ee

  \end{document}